\documentclass[a4paper, fleqn,usenatbib]{mnras}

\usepackage{newtxtext,newtxmath}

\usepackage[T1]{fontenc}
\usepackage{ae,aecompl}

\usepackage{graphicx}	
\usepackage{amsmath}	
\usepackage{amssymb}	

\title[Star-Formation Properties of AGN: BAT vs EAGLE]{The Star-Formation Properties of the Observed and Simulated AGN Universe: BAT vs EAGLE}

\author[Thomas. M. Jackson et al.]{
Thomas M. Jackson$^{1,2}$,\thanks{E-mail: Thomas.jackson@uni-heidelberg.de}
D. J. Rosario$^{2}$,
D. M. Alexander$^{2}$,
J. Scholtz$^{2}$,
\newauthor{Stuart McAlpine$^{3}$, R. G. Bower$^{4}$}
\\
$^{1}$Astronomisches Rechen-Institut, M{\"o}nchhofstr. 12-14, D-69120 Heidelberg, Germany\\
$^{2}$Centre for Extragalactic Astronomy, Department of Physics, Durham University, South Road, Durham, DH1 3LE, UK\\
$^{3}$ Department of Physics, University of Helsinki, Gustaf H{\"a}llstr{\"o}min katu 2a, P.O. Box 64, FI-00014 University of Helsinki, Finland\\
$^{4}$Institute for Computational Cosmology, Department of Physics, Durham University, South Road, Durham, DH1 3LE, UK}

\date{Accepted 2020 August 11. Received 2020 August 11; in original form 2020 April 7}

\pubyear{2020}

\begin{document}
\label{firstpage}
\pagerange{\pageref{firstpage}--\pageref{lastpage}}
\maketitle

\begin{abstract}
In this paper we present data from 72 low redshift, hard X-ray selected AGN taken from the {\it Swift}-BAT 58 month catalogue. We utilise spectral energy distribution fitting to the optical to IR photometry in order to estimate host galaxy properties. We compare this observational sample to a volume and flux matched sample of AGN from the EAGLE hydrodynamical simulations in order to verify how accurately the simulations can reproduce observed AGN host galaxy properties. After correcting for the known +0.2 dex offset in the SFRs between EAGLE and previous observations, we find agreement in the SFR and X-ray luminosity distributions; however we find that the stellar masses in EAGLE are $0.2 - 0.4$ dex greater than the observational sample, which consequently leads to lower sSFRs. We compare these results to our previous study at high redshift, finding agreement in both the observations and simulations, whereby the widths of sSFR distributions are similar ($\sim0.4-0.6$ dex) and the median of the SFR distributions lie below the star forming main sequence by $\sim0.3-0.5$ dex across all samples. We also use EAGLE to select a sample of AGN host galaxies at high and low redshift and follow their characteristic evolution from $z=8$ to $z=0$. We find similar behaviour between these two samples, whereby star formation is quenched when the black hole goes through its phase of most rapid growth. Utilising EAGLE we find that 23\% of AGN selected at $z\sim0$ are also AGN at high redshift, and that their host galaxies are among the most massive objects in the simulation. Overall we find EAGLE reproduces the observations well, with some minor inconsistencies ($\sim$ 0.2 dex in stellar masses and $\sim$ 0.4 dex in sSFRs).
\end{abstract}




\section{Introduction}
\label{sec:Introduction}

It has long been known that certain properties of galaxies are related to the mass of their central supermassive black hole (SMBH). Examples include the tight correlation found between the black hole mass and the bulge mass/stellar velocity dispersion \citep{1998AJ....115.2285M, 2004ApJ...604L..89H, 2013ARA&A..51..511K, 2013ApJ...764..184M} or the similar evolution of the average cosmic star formation rate (SFR) and average cosmic black hole accretion rate (BHAR) \citep{1998MNRAS.293L..49B, 2010MNRAS.401.2531A, 2015MNRAS.451.1892A}. 

Both theoretical models \citep{1998A&A...331L...1S} and cosmological simulations \citep{2006MNRAS.370..645B, 2006MNRAS.367..864C, 2017MNRAS.465...32B} have been used to support the idea that energy originating from an accreting SMBH (an Active Galactic Nucleus or AGN) during its growth phases can be injected back into the interstellar medium, thereby regulating the inflow of gas and star formation. This process is called AGN feedback, and is postulated to be responsible for the relations we observe. 

Indeed most successful cosmological simulations require AGN feedback in order to reproduce multiple properties of the observed universe, indicating that AGN feedback may play a significant role in the evolution of the galaxy population. Some of these properties include the present day stellar mass function \citep{2006MNRAS.370..645B, 2006MNRAS.367..864C}, the colour bi-modality of galaxies in the colour magnitude plane, the black hole - spheroid relationship and the metallicity of the intergalactic medium \citep{2012NewAR..56...93A, 2012ARA&A..50..455F, 2017NatAs...1E.165H}. 

Numerous observational studies have attempted to find direct evidence of AGN feedback \citep[e.g.][]{2012A&A...545A..45R, 2012ApJ...760L..15H}, however understanding and separating feedback from fuelling is a complex proposition since AGN hosts span a range of accretion rates, BH masses, galaxy properties, etc. Star formation shares the same fuel as a luminous accreting SMBH - cold gas. Star formation is also the primary process by which galaxies grow in stellar mass (discounting mergers or other similar processes), making star formation a prime candidate in the search for imprints of AGN feedback.   

Many studies agree on a flat trend between the average host galaxy SFR as a function of AGN luminosity for moderate luminosity AGN (10$^{43}$ erg s$^{-1}$ < L$_{\mathrm{bol}}$ < 10$^{45}$ erg s$^{-1}$) \citep{2012A&A...545A..45R, 2015MNRAS.453..591S, 2017MNRAS.466.3161S}, where L$_{\mathrm{bol}}$ is the AGN bolometric luminosity. Although this flat relation may be an unexpected result, \citet{2014ApJ...782....9H} argue that the reason for this behaviour is the different timescales upon which star formation and AGN luminosity vary. Significant differences in the luminosity, driven by changes in the average accretion rate, are expected over a timescale on the order of a Myr \citep{2018MNRAS.476L..34S}. The luminosity due to star formation varies on the order of galaxy dynamical timescales \citep[See][and references therein]{2012ARA&A..50..531K}. These differences in timescale wash out any expected correlation. Studies using simulations support this concept \citep{2017MNRAS.468.3395M}. 

Cosmological hydrodynamical simulations can provide a solution to these problems. They firstly apply a self-consistent treatment of both dark matter physics and baryonic physics. They secondly have the advantage of being able to track the cosmological growth of a large enough number of model galaxies for the statistical sampling of diverse behaviour, while still allowing for the tracking of gas and related processes on relatively small scales \citep[e.g.][]{2012MNRAS.419.3200H}. For example the EAGLE simulations (IllustrisTNG50, IllustrisTNG300) contains $\sim 15,000$ galaxies ($\sim 2,000$, $\sim 100,000$) with a stellar mass of M$_{\mathrm{Stellar}}$ > 10$^{9}$ M$_{\odot}$ while still retaining a baryonic mass resolution, m$_{\mathrm{baryon}}$ of $1.81 \times 10^6$ M$_{\odot}$ ($8.5 \times 10^4$ M$_{\odot}$, $1.4 \times 10^7$ M$_{\odot}$) \citep{2015MNRAS.446..521S, 2018MNRAS.473.4077P}. This means that each galaxy can be represented by between 10$^2$ to 10$^7$ particles and allows spatial resolutions down to $\sim 70$pc \citep{2019arXiv190205553P}. This is a significant advantage when trying to disentangle the various processes which drive star formation and AGN fuelling and feedback while still retaining statistical power. 

AGN feedback in hydrodynamical simulations is implemented primarily to quench star formation in high mass galaxies, thereby reducing the maximum stellar mass a galaxy may assemble \citep{2006MNRAS.370..645B}. Depending on the simulation, the feedback can be delivered thermally \citep{2015MNRAS.446..521S}, mechanically \citep{2019MNRAS.486.2827D} or using a mixture of both \citep{2016MNRAS.463.3948D, 2018MNRAS.479.4056W}. The level of AGN feedback is usually calibrated with a small number of other parameters within the sub-grid physics representing the physical processes below the resolution scale in order to reproduce broad galaxy properties (see Section~\ref{sec:sims} for more detail).

In order to find the imprints of AGN feedback on star formation, \citet{2018MNRAS.475.1288S} compared EAGLE models with and without AGN feedback. They found that AGN feedback quenched star formation in higher mass galaxies (M$_{\mathrm{Stellar}}$ > 10$^{10}$ M$_\odot$), thereby broadening the specific Star Formation Rate (sSFR) distributions of the galaxy population. They also found good agreement between the sSFR distributions of their observational sample at high redshift ($1.5 < z < 3.2$) and a comparison sample obtained by applying the same redshift, stellar mass and X-ray luminosity cuts to the EAGLE models incorporating AGN feedback. 

In this paper, we compare the EAGLE hydrodynamical simulations to a well-defined set of AGN from the {\it Swift}-BAT Ultra-hard X-ray survey. We apply matching criteria and observational biases to AGN and galaxies in the simulation in order to meaningfully compare the host galaxy properties to our observational sample of low redshift AGN (distance $\lesssim$ 70 Mpc). Nearby AGN are generally well studied and offer more accurate constraints on their host galaxy properties such as SFR, stellar mass and AGN luminosity in comparison to high redshift AGN. This provides a comprehensive and important validation exercise for the treatment of accretion and feedback in EAGLE by comparing if trends and various host galaxy property distributions are reproduced. 

The significant redshift difference between this research and that of \citet{2018MNRAS.475.1288S} allows us to firstly check for consistencies (or lack thereof) in both the observations and simulations and secondly to investigate how the epoch of selection of the AGN sample may affect the characteristic host galaxy properties and fueling processes as a function of redshift. We investigate this by comparing the cosmic evolution of AGN host galaxy properties within EAGLE of two AGN samples, one selected at low redshift and one at high redshift. 

In Section~\ref{sec:Sample} we present the observational sample and methods applied to the data. In Section~\ref{sec:sims} we briefly outline the simulations and describe the methods applied to our simulated comparison samples. In Section~\ref{sec:Results} we present our results of the comparison of observational and simulated data and the exploration of the simulation. In Section~\ref{sec:disc} we discuss the context of our results and in Section~\ref{sec:Conc} we summarise our work. In our work we assume the cosmological parameters from the Planck mission \citep{2015A&A...582A..31P}, whereby $\Omega_\Lambda = 0.693$, $\Omega_m = 0.307$, $\Omega_b = 0.04825$ and $H_0 = 67.77$ km s$^{-1}$.

\section{Observational Data and Methods}
\label{sec:Sample}

In this section we describe the observational data and methods used in this research. We present the X-ray selection of the AGN sample (see Section~\ref{subsec:x-ray selection}), the collection of associated multi-wavelength photometry (see Section~\ref{subsec:IR and Optical}), and the SED fitting procedure applied to the photometry in order to estimate the host galaxy SFRs and stellar masses (see Section~\ref{subsec:SFR}).

\subsection{X-ray selection of AGN}
\label{subsec:x-ray selection}

\begin{figure*}
	\centering
	\includegraphics[width=0.9\linewidth]{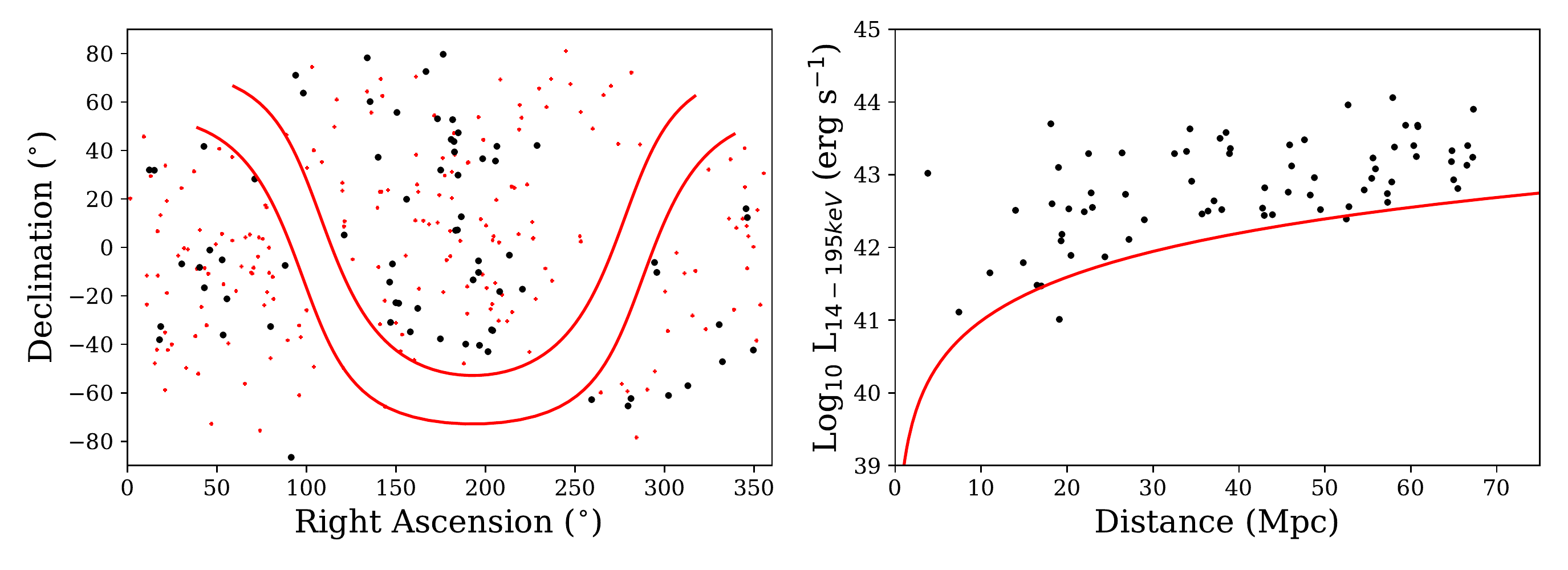}
	\caption{The left hand panel shows the distributions of our BAT AGN sample (black points) and the parent sample from the {\it Swift}-BAT 58-month catalogue (red crosses) in RA and DEC. The 10$^{\circ}$ limits from the central Milky Way plane are shown by the solid red lines. The right panel shows the distribution in redshift versus X-ray luminosity (L$_{14 - 195 \mathrm{keV}}$) for the BAT AGN sample with the approximate flux limit represented by the solid red line.}
	\label{fig:sample}
\end{figure*}  

AGN can be identified in many different bands of the electromagnetic (EM) spectrum such as the optical, Infra-red (IR) or radio \citep{2003MNRAS.346.1055K, 2009ApJ...699L..43S, 2013ApJS..208...24L}. Selection of AGN in the ultra hard X-ray band ($E \gtrsim$ 10 keV), however, can be advantageous. Firstly, there are relatively few sources other than AGN which produce Ultra hard X-ray emission, reducing the chance of contamination in both sample selection and luminosity measurements \citep{2009MNRAS.398..333H}. Secondly, absorption of Ultra hard X-rays only occurs through high column densities of gas along the line of sight \citep[N$_{\mathrm{H}} \gtrsim 10^{24}$ cm$^{-2}$, e.g.][]{2000ApJ...542..914W}. Other parts of the EM spectrum such as the Ultraviolet (UV) or optical may be blocked at much lower column densities when gas and dust are present \citep[see][and references therein]{2018ARA&A..56..625H}.

Selection in the Ultra hard X-ray regime therefore provides a relatively unbiased sample of AGN compared to IR or radio selection \citep{2000ApJ...528..637B, 2004ASSL..308...53M}. It can also lead to the discovery of new, highly obscured AGN \citep{2010ApJ...716L.125K, 2013ApJS..207...19B, 2017ApJ...846...20L}. 

We use the 58 month catalogue\footnote{\url{https://swift.gsfc.nasa.gov/results/bs58mon/}} of sources from the NASA {\it Swift} satellite's Burst Alert Telescope (hereafter {\it Swift}-BAT)\footnote{\url{https://swift.gsfc.nasa.gov/}}, which operates in the Ultra-hard X-ray band (14 - 195 keV), to select our sources.

In order to reduce the contamination that Galactic sources may have on any targets in both the the optical and IR photometry described in Section~\ref{subsec:IR and Optical}, objects within $\pm$ 10$^{\circ}$ of the Galactic plane were removed. In order to reduce statistical biases that can emerge due to volume driven effects (the larger the volume the more statistically likely it is to have objects with extreme luminosities, stellar masses, SFRs etc.) we matched the observational volume to the EAGLE simulation volume (in this study we use the 100 cMpc box size, described in Section~\ref{subsec:EAGLE}) by applying a cut-off in luminosity distance of 67.8 Mpc to the parent sample in the {\it Swift}-BAT 58 month catalogue. 

The final observational sample contains 72 objects, which are hereafter referred to as our BAT AGN sample. These differences in sources can be seen in the left hand panel of Figure~\ref{fig:sample}, where the black points represent our BAT AGN sample and the red points represent sources cut from the 58 month catalogue (after removing those sources within the Galactic plane). 

To obtain the most up-to-date and uniformly assessed redshifts and X-ray luminosities, the data were matched to the BAT AGN Spectroscopic Survey  \citep[BASS\footnote{\url{https://www.bass-survey.com/}},][]{2017ApJ...850...74K}, an optical and IR spectroscopic survey of {\it Swift}-BAT sources, estimating redshifts from the optical spectra and collecting the X-ray luminosities from the combined processing of Swift/BAT and softer X-ray datasets from XMM-Newton, {\it Chandra}, etc. \citep{2017ApJS..233...17R}. The distribution in X-ray luminosity and redshift of the BAT AGN sample can be seen in the right hand panel of Figure~\ref{fig:sample}. A table of the sample and its estimated properties can be found in Appendix~\ref{app: BAT AGN}.

\subsection{IR and optical counterpart data}
\label{subsec:IR and Optical}

Host galaxy properties such as stellar mass and SFR can be estimated via Spectral Energy Distribution (SED) fitting. This procedure is described in Section~\ref{subsec:SFR}. In order to fit an SED to an AGN and its host galaxy to estimate these properties, multiple photometric data points spanning the optical and IR are needed. Due to the all-sky nature of the {\it Swift}-BAT sample, data from various surveys were collected.

To obtain uniform counterpart photometry in the Near and Mid Infra-red (NIR and MIR) for as many of our observed AGN as possible, we used the all-sky datasets of the Wide-field Infra-red Survey Explorer ({\it WISE})\footnote{\url{https://www.nasa.gov/mission_pages/WISE/main/index.html}} \citep{2010AJ....140.1868W} and the 2 Micron All Sky Survey (2MASS)\footnote{\url{https://www.ipac.caltech.edu/2mass/}} \citep{2006AJ....131.1163S}. We compiled MIR photometry from the online database of the {\it WISE} mission, served on the NASA/IPAC website, spanning four bands at 3.5, 4.6, 12 \& 22 $\mu$m. Since our targets are well-resolved nearby galaxies, we used the GMAG photometric measurements in all cases. These are photometry using apertures scaled to the 2MASS Extended Source Catalog (XSC) \citep{2000AJ....119.2498J} size measurements, with a correction for the Point Spread Function of the {\it WISE} images. We converted {\it WISE} magnitudes to fluxes using the conversions from the online {\it WISE} Explanatory Supplement. The all-sky 2MASS XSC provides photometry in 3 NIR bands (J, H, K$_{s}$) for all the BAT AGN. We used the isophotal curve-of-growth magnitudes in this work, and employ standard conversions from 2MASS magnitudes to fluxes from the online 2MASS Explanatory Supplement. 

To provide better constraints on the star-forming component of each AGN (as explained in Section~\ref{subsec:SFR}), Far Infra-red (FIR) photometric data were added. We obtained {\it Herschel} photometry for the BAT AGN from the data release described in \citet{2017MNRAS.466.3161S}, based on the analysis presented in \citet{2014ApJ...794..152M} (PACS) and \citet{2016MNRAS.456.3335S} (SPIRE). We used their measurements for the integrated photometry of the sources, covering 5 bands: PACS 70 \& 160 $\mu$m and SPIRE 250, 350 \& 500 $\mu$m.

The collection of uniform and accurate optical photometry for the BAT AGN sample is a more complex process than for the IR. The NASA Extragalactic Database (NED) supplies a heterogeneous dataset of photometry from the literature. We obtained B band photographic photometry of a subset of galaxies from the NED version of the RC3 catalog \citep{1991MNRAS.249P..28D, 1994AJ....108.2128C}. The SDSS \citep{2000AJ....120.1579Y} and PanSTARRS \citep{2016arXiv161205560C} optical imaging surveys yielded magnitudes in 6 bands from $u$--$y$. From the photometric catalogs of these surveys, we used the CMODEL magnitudes, which combine two different light profile models of extended sources to produce best-effort integrated photometric measurements. Where possible, we replace the SDSS database magnitudes with those from the NASA Sloan Atlas (NSA), a value-added resource for nearby galaxies with more accurate photometry for extended sources. We also used, where available, HyperLEDA measurements in five bands (UBVRI) which are homogenised curve-of-growth photometry as presented in \citet{1998A&AS..128..299P}. Finally, for a small number of galaxies, curve-of-growth galaxy photometry in the BVRI bands is available from the Carnegie-Irvine Galaxy survey (CGS; \citet{2011ApJS..197...21H}). 

To ensure the highest quality integrated photometry that is estimated uniformly, we restricted our optical photometry to only one of these datasets for each galaxy, taking the following priority in decreasing order: The Carnegie-Irvine galaxy survey\footnote{\url{https://cgs.obs.carnegiescience.edu/CGS/Home.html}}, the HyperLEDA extragalactic survey\footnote{\url{http://leda.univ-lyon1.fr/}}, the NSA\footnote{\url{http://www.nsatlas.org/}}, SDSS\footnote{\url{https://www.sdss.org/}}, Pan-STARRs\footnote{\url{https://panstarrs.stsci.edu/}} and finally RC3 from NED\footnote{\url{https://ned.ipac.caltech.edu/}}.

\subsection{SED fitting of the BAT AGN using FortesFit}
\label{subsec:SFR}

SFRs can be estimated using various techniques; for example conversions from H$\alpha$, total UV light and radio emission from supernova to name a few \citep[see][for a summary and references therein]{2012ARA&A..50..531K}. Most of these methods can be significantly contaminated by emission from the AGN itself. The UV can be dominated by emission from the accretion disk, an effect especially prominent for unobscured AGN, estimations of the SFR in the radio can be contaminated by non-thermal radiative processes such as synchrotron radiation emitted from relativistic jets and lobes and/or the central regions and H$\alpha$ can be contaminated due to photoionisation from the AGN. 

Star formation contributes strongly at FIR and sub-mm wavelengths. This is due to the re-processing of UV or optical light from young stars via the dust rich clouds that typically surround star forming regions. Similarly, the AGN dusty torus also emits in the IR, however due to the proximity of the torus to the accretion disk the temperature of the dust is higher than that of the dust around star forming regions \citep[see][and references herein]{2007ApJ...666..806N}. Therefore the emission peaks in the MIR, dropping off sharply as we reach the FIR, where re-emitted light from star formation dominates \citep{2011MNRAS.414.1082M, 2012A&A...545A..45R}.

In order to estimate SFRs and stellar masses, we performed template fitting to the photometry ranging from the UV to FIR that comprises the overall SEDs of our sources. We used the FortesFit\footnote{\url{https://github.com/vikalibrate/FortesFit}} SED template fitting routine for this task. FortesFit is a python-based fully Bayesian fitting routine. It contains a library of parameterised model SEDs that work additively in order to try and fit the full SED to the data points provided. Priors of certain parameters can be provided, such as the AGN type classification, in order to better constrain the expected contribution of various components, thereby minimising the chance of mis-fitting. The fitting routine then provides posterior probability distributions for various values such as the AGN IR luminosity component, stellar mass and the integrated luminosity in the IR due to star formation. For further details of the SED modelling, refer to \citet{2020MNRAS.492.3194S} 

To convert the integrated IR luminosity component (in erg s$^{-1}$, integrated between 8-1000 $\mu$m) due to star formation into the SFR, we use Equation~\ref{eq:1} below \citep[equation 6 from][see the paper for further details of the full assumptions and models]{2016MNRAS.457.2703R}. We also assume an intrinsic uncertainty of 0.25 dex in this calibration \citep{2016MNRAS.457.2703R}, which we take into account in further analysis.

\begin{equation}
\label{eq:1}
SFR = \dfrac{L_{\mathrm{IR, SF}}}{4.48 \times 10^{43}} $ (M$_\odot$ yr$^{-1}$)$
\end{equation}

These estimates and other data for our BAT AGN can be found in Appendix~\ref{app: BAT AGN}.

\section{Simulation data and methods}
\label{sec:sims}

In this section we provide a brief outline of the simulations (see Section~\ref{subsec:EAGLE}). We then describe processes such as the initial constraints used to attain the comparison data set and the calculation of some of the relevant host galaxy properties (see Section~\ref{subsec:sim_sample}). Finally, we describe the use of the merger trees system adopted to track the characteristic host galaxy property evolution of the AGN comparison sample and the simulated AGN comparison sample in \citet[][see Section~\ref{subsec:SFH}]{2018MNRAS.475.1288S}. 

\subsection{EAGLE}
\label{subsec:EAGLE}

The Evolution and Assembly of GaLaxies and their Environments (EAGLE) is a suite of hydrodynamical simulations \footnote{\url{http://icc.dur.ac.uk/Eagle/}} \citep{2015MNRAS.450.1937C, 2015MNRAS.446..521S}. The simulations use an improved version of the GADGET-3 Smoothed Particle Hydrodynamics (SPH) code \citep{2005MNRAS.364.1105S}, with the cosmological parameters from the Planck mission \citep{2015A&A...582A..31P} implemented, whereby $\Omega_\Lambda = 0.693$, $\Omega_m = 0.307$, $\Omega_b = 0.04825$ and $H_0 = 67.77$ km s$^{-1}$ which we also adopt for any calculations used in this paper. To account for all processes which operate below the numerical resolution limit, the simulations employ sub-grid physics prescriptions relying only on local hydrodynamic properties. This accounts for radiative cooling and photo-ionisation heating \citep{2009MNRAS.393...99W}, star formation \citep*{2008MNRAS.383.1210S}, stellar mass loss \citep{2009MNRAS.399..574W} and stellar feedback \citep*{2012MNRAS.426..140D} and black hole growth and feedback. These parameters are then calibrated to reproduce the stellar mass function, the black hole mass - bulge mass relation and the galaxy stellar mass - size relation at $z = 0.1$. A full description of the suite of simulations and calibrations can be found in \citet{2015MNRAS.446..521S} and \citet{2015MNRAS.450.1937C}.

For the purposes of this work, star formation and black hole accretion are the most relevant sub-grid physical processes in EAGLE. Star formation is modelled according to \citet{2008MNRAS.383.1210S} as a stochastic process based on the pressure dependent Kennicutt-Schmidt law with a \citet{2003PASP..115..763C} IMF. Black holes are seeded in the densest gas particle in every dark matter halo greater than 1.48 $\times$ 10$^{10}$ M$_{\odot}$. The black holes then grow through accretion of other gas particles. To prevent over-cooling, the energy released from accretion, assuming a radiative efficiency $\epsilon = 0.1$, is not released immediately into the gas particles surrounding the black hole, but stored as an energy reservoir for a period of time given by an injection probability distribution.

EAGLE has been shown to also reproduce many properties of galaxies in the local and higher redshift universe without requiring specific tuning of the simulation parameters, such as galaxy colours, the Tully-Fisher relation, the evolution of the cosmic average SFR, the passive galaxy fraction, rotation curves and metallicities \citep{2015MNRAS.451.1247S, 2015MNRAS.452.3815L, 2015MNRAS.452.2879T, 2015MNRAS.450.4486F, 2015MNRAS.446..521S}. Similar to these studies, our comparison serves as a somewhat independent test of the EAGLE model, as the simulations have not been calibrated specifically to reproduce the star formation properties of AGN host galaxies. 

The simulation also has various simulation box sizes, resolutions and physical models \citep{2016MNRAS.462..190R} such as higher AGN heating temperature and increased black hole viscosity. Further details of these variations can be found on the EAGLE database \citep{2016A&C....15...72M}. The standard model referred to hereafter, incorporates AGN feedback and standard values for multiple parameters which can be found in \citet{2015MNRAS.450.1937C} as RefL0100N1504. 

\subsection{Simulated comparison sample}
\label{subsec:sim_sample}

To obtain the properties of galaxies within EAGLE we queried the public database using the SQL interface \citep{2016A&C....15...72M}. The key properties in low mass haloes are increasingly inaccurate due to the limitations of numerical resolution, so we only considered galaxies with stellar mass M$_{\mathrm{Stellar}} > 10^8 $M$_\odot$ (i.e the number of particles $N \gtrsim 100$). To ensure uniform measurements of properties such as SFR or BHAR across the simulated galaxy sample we also applied an aperture of 30 kpc. This is the same aperture as used in previous studies and defines the maximal distance of particles directly associated with a galaxy \citep{2016A&C....15...72M}. 

Stellar mass, black hole mass and halo mass are directly available from the database. EAGLE also reproduces the shape of the cosmic evolution of the average SFR well, however \citet{2015MNRAS.450.4486F} found that EAGLE predicts a $\simeq$ 0.2 dex lower average SFR over all epochs, possibly the result of calibration uncertainties. Therefore, following \citet{2017MNRAS.468.3395M}, we also applied a + 0.2 dex offset to all the SFRs from the EAGLE database \citep[see also][for more discussion]{2018MNRAS.475.1288S}. We make it clear at this point that by applying an offset in the SFR measurements, we are not representing the true star formation history (SFH) of a galaxy and thereby not treating the stellar mass measurements self-consistently. We leave a fuller discussion of this, however, to Section~\ref{sec:disc}.

Most of the galaxies in EAGLE, however, are inactive galaxies or AGN that would be too faint to be detected by {\it Swift}-BAT. These objects are usually lower in SFR and/or stellar mass and hence introduce a bias into our results. Stellar mass and/or X-ray luminosity matched samples may provide good comparison samples, however in order to investigate to what degree EAGLE reproduces the local universe we needed to apply the same selection to EAGLE as that of the {\it Swift}-BAT sample, namely the sensitivity or hard X-ray flux limit of the instrument. 

To calculate the hard X-ray luminosity, we followed \citet{2017MNRAS.468.3395M}, converting the BHAR ($\dot{m}$) given by the database into bolometric luminosity using the following conversion: L$_{\mathrm{bol}} = \epsilon \dot{m}c^2$, assuming a radiative efficiency of $\epsilon = 0.1$ and where L$_{\mathrm{bol}}$ is the bolometric luminosity of the AGN. We then converted the bolometric luminosity to hard X-ray luminosity with the prescription given by \citet{2017MNRAS.470..800T}, whereby L$_{\mathrm{bol}} =$ 8.5 $\times$ L$_{14 - 195 \mathrm{keV}}$. We also accounted for extinction caused by obscuration of the AGN as EAGLE does not contain a prescription for the high levels of extinction that may occur near an AGN. 

\citet{2015ApJ...815L..13R} estimated the fraction of AGN with various column densities, ranging from N$_{\mathrm{H}}$ = 10$^{20}$ cm$^{-2}$ - 10$^{25}$ cm$^{-2}$ \citep[figure 4 of][]{2015ApJ...815L..13R} and the fraction of intrinsic X-ray flux which is absorbed along the line of sight as a function of column density \citep[figure 1 of][]{2015ApJ...815L..13R}. We used these distributions and randomly assigned a column density to each AGN according to the probability fraction. We then adjusted the X-ray luminosity value according to the extinction expected from this assigned column density. This process reduces the simulated AGN counts from an average of 159 sources to 132 sources ($\sim$ 15 per cent reduction), once the flux matching process is applied as described below.

To convert the X-ray luminosity to X-ray flux we calculated the distances from each galaxy to the centre of the simulation box. We then used these distances in conjunction with the converted and corrected X-ray luminosities in order to find the expected X-ray flux. {\it Swift}-BAT does not have homogeneous coverage of the sky (i.e the flux sensitivity varies for certain areas of the sky). 50\% of the sky is covered down to a flux limit of 1.1 $\times$ 10$^{-11}$ erg s$^{-1}$ cm$^{-2}$, 40\% of the remaining sky down to 1.5 $\times$ 10$^{-11}$ erg s$^{-1}$ cm$^{-2}$ and the remaining 10\% to 1.7 $\times$ 10$^{-11}$ erg s$^{-1}$ cm$^{-2}$. We therefore used a random number generator to assign one of these flux limits to each galaxy according to the relative probability. If the estimated X-ray flux was above the randomly assigned flux limit, then the galaxy was retained. If the estimated X-ray flux was below the assigned limit, the galaxy was discarded. This process yields what we hereafter call a single realisation of the EAGLE AGN. We then repeat this process 10,000 times to fully utilise the statistical power of EAGLE (results seen in Section~\ref{res: sfr}), hereafter the EAGLE AGN sample. Overall we find, on average, 132 EAGLE AGN per realisation compared to our 72 BAT AGN. We discuss this factor $\sim 2$ discrepancy further, as well as other bolometric to X-ray conversions, in Section~\ref{sec:disc}.

\begin{figure}
	\centering
	\includegraphics[width=0.9\linewidth]{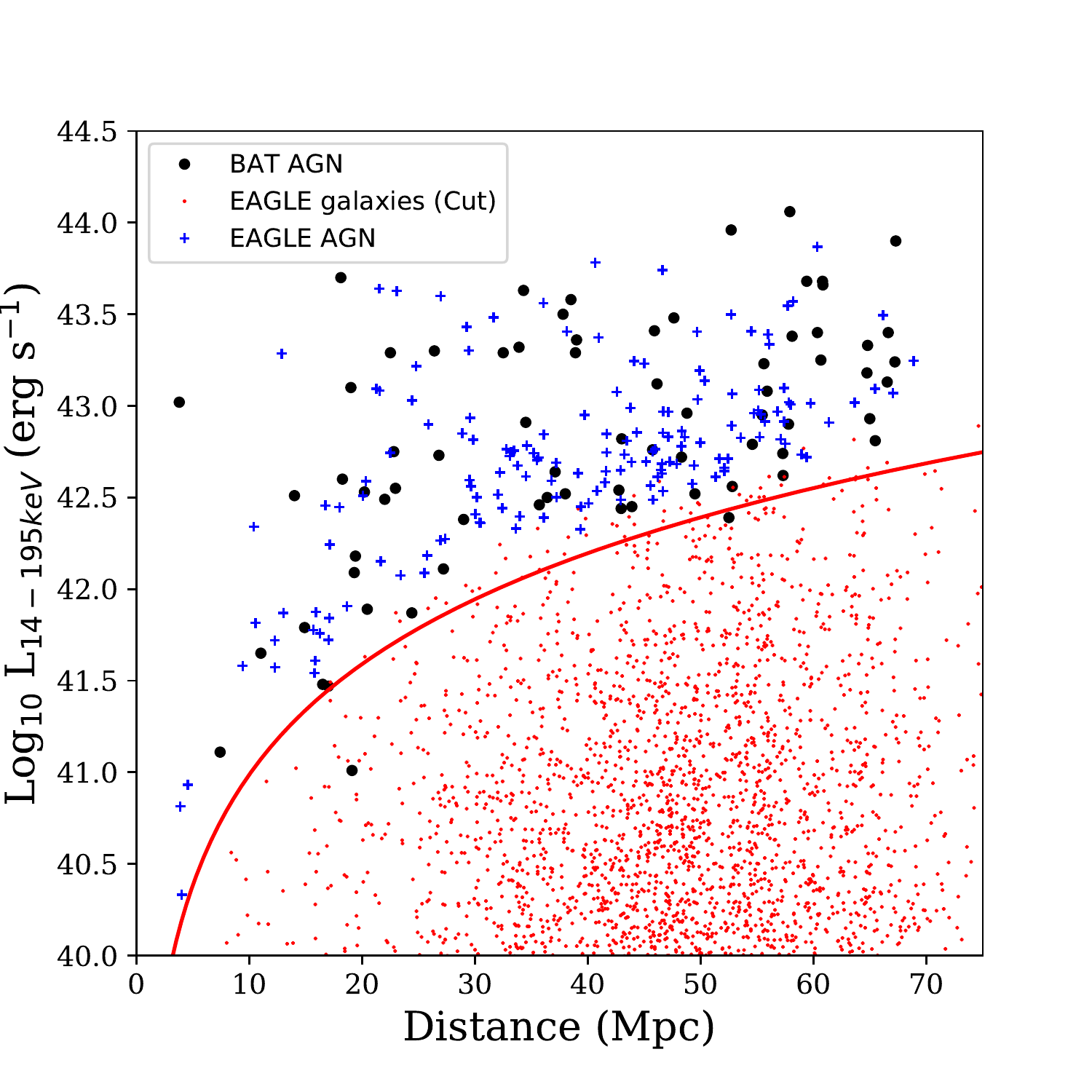}
	\caption{The distribution in redshift and X-ray luminosity of our BAT AGN (black points), our EAGLE AGN (blue points) and those galaxies discarded by our flux matching process (red points). The approximate sensitivity of Swift-BAT instrument is represented by the solid red line.}
	\label{fig:eag_agn}
\end{figure}  

The results of this process can be seen in Figure~\ref{fig:eag_agn}. The black points represent our BAT AGN sample, the blue points represent the AGN retained for a single realisation of the comparison sample (hereafter a single realisation of the EAGLE AGN) and the red points represent the galaxies in EAGLE that were discarded from the flux matching process. The solid red line shows the approximate flux limit of {\it Swift}-BAT. To investigate how much variance our process of flux correction and matching introduces, and to fully utilise the full statistical power of EAGLE, we also run the entire flux matching process 10,000 times. 

\subsection{Use of the merger trees within EAGLE}
\label{subsec:SFH}

One advantage of simulations is that each galaxy and its properties can be followed from the present day up to high redshift in various snapshots. This allows the exploration of the evolution of a population of galaxies over time and the properties of their progenitors. We explored the host galaxy evolution of a single realisation of our EAGLE AGN compared to the AGN selected at high redshift from EAGLE in the work of \citet{2018MNRAS.475.1288S}. This allows us to investigate if AGN selected at the peak of cosmic AGN and star forming activity ($z \sim 2$), are inherently different, i.e. if key growth processes, as modelled by the simulation, differ at these two epochs.

The individual histories of galaxies in EAGLE can be traced over 28 snapshots from $z=0$ to $z = 20$ using the merger tree system. To evaluate the entire assembly history of a galaxy at $z = 0$, the paths between the present day galaxy and all of its progenitors can be traced, creating an entire tree. Alternatively, to simplify the evolutionary history, the main branch of the galaxy can be traced from the present day galaxy to the galaxy with the TopLeafID, i.e. the path of the 'main' or most massive progenitor at each snapshot. This is a more meaningful comparison as the main branch is where the majority of the overall growth of both stellar material and SMBH mass, capturing the gross statistical trends.

A major merger may greatly affect the properties of a particular simulated galaxy between any two snapshots, i.e. a sudden and significant increase in stellar mass and/or star formation rate. This could be a potential source for biases when comparing the average galaxy properties of two different snapshots or epochs. The simulated major merger rate is low however, with half of all galaxies in the simulations with stellar mass $10^{9.5} <$ M$_{\odot}$ $< 10^{10.5}$ ($10^{10.5} <$ M$_{\odot}$ $< 10^{11}$, $10^{11} <$ M$_{\odot}$ $< 10^{12}$) having not undergone a major merger since $z \sim 4$ \citep[$z \sim 3$, $z \sim 1$, ][]{2017MNRAS.464.1659Q}. This means that when we compare the evolution of characteristic host galaxy properties of an entire AGN sample, the biases that any one major merger may contribute is likely to be washed out.

The EAGLE AGN sample from \citet{2018MNRAS.475.1288S} was selected to have a stellar mass M$_{\mathrm{Stellar}} >$ 10$^{10}$ M$_\odot$, an X-ray luminosity L$_{\mathrm{2 - 10 keV}}$ > 10$^{43}$ - 10$^{45}$ and a redshift range of $1.4 < z < 3.6$, yielding 292 AGN (hereafter S18-EAGLE-AGN). We compared the evolution of the host galaxy properties of the S18-EAGLE-AGN to a single realisation of the EAGLE AGN; we note that although we only selected one realisation, the overall trends are not significantly dependent on different realisations.  

To allow the S18-EAGLE-AGN sample to be tracked back far enough in cosmic time before the epoch of selection, we chose a redshift cut-off of $z = 8$. This redshift cut-off also has the advantage that the majority of AGN host galaxies in both samples still have a sufficient number of stellar particles to avoid the statistical inaccuracies introduced by low counts. We queried the SFRs, stellar masses, black hole masses, and halo masses at each snapshot between $0 < z < 8$. In order to quantify the characteristic host galaxy properties at each snapshot, the median values of each quantity and in each redshift snapshot were calculated and the 16$^{\mathrm{th}}$ and 84$^{\mathrm{th}}$ percentiles of the distributions calculated. Finally, to investigate under what conditions AGN feedback plays a role in quenching star formation within EAGLE (i.e. as indicated by a decrease in star formation relative to average star forming galaxies) we also compared our EAGLE AGN sample and the S18-EAGLE-AGN to their respective star forming main sequences (hereafter SF-MS).
	
For the SF-MS we used the prescription from \citet{2015A&A...575A..74S}. This study utilised a mixture of photometry from previous surveys and stacked galaxy imaging from $Herschel$, spanning multiple stellar mass bins (between 10$^{9.5}$ < M$_{\mathrm{Stellar}} < 10^{11.5}$ M$_\odot$) and redshifts (0.3 $< z <$ 5). They then selected star forming galaxies using $UVJ$ colour selection and fitted SEDs from the UV to the FIR in order to estimate galaxy properties. These estimates were then utilised to construct an average SF-MS as a function of stellar mass and redshift. 

We also found that 31 galaxies within the EAGLE simulations are contained in both samples  \citep[i.e. 23\% of our EAGLE AGN galaxies selected at $z = 0$ are selected as AGN at $1.4 < z < 3.6$, according to the selection criteria of][]{2018MNRAS.475.1288S}. In order to further investigate this sub-population of galaxies which display significant re-current AGN episodes at both high and low redshift, we also tracked their host galaxy property evolution throughout cosmic time in the same manner as the EAGLE AGN and the S18-EAGLE-AGN.

\section{Results}
\label{sec:Results}

In this section we present the results from our analyses. We show a comparison of the SFR versus X-ray luminosity trends of the BAT AGN against a single realisation of the EAGLE AGN sample and the respective distributions (see Section~\ref{res: sfr}). We then show the comparison of the distributions in SFR, stellar mass, sSFR and X-ray luminosity of the BAT AGN and the EAGLE AGN samples and the position of each sample with respect to the SF-MS (see Section~\ref{res: ssfr}). We also present the results from a Monte-Carlo (MC) analysis (see Section~\ref{res: MC}) and finally our comparison of the evolution over redshift of average host galaxy properties of two different samples selected at different redshifts (plus the evolution of those galaxies contained in both samples, see Section~\ref{res: trees}).

\subsection{Comparison of SFR versus X-ray luminosity}
\label{res: sfr}

\begin{figure}
	\centering
	\includegraphics[width=1.1\linewidth]{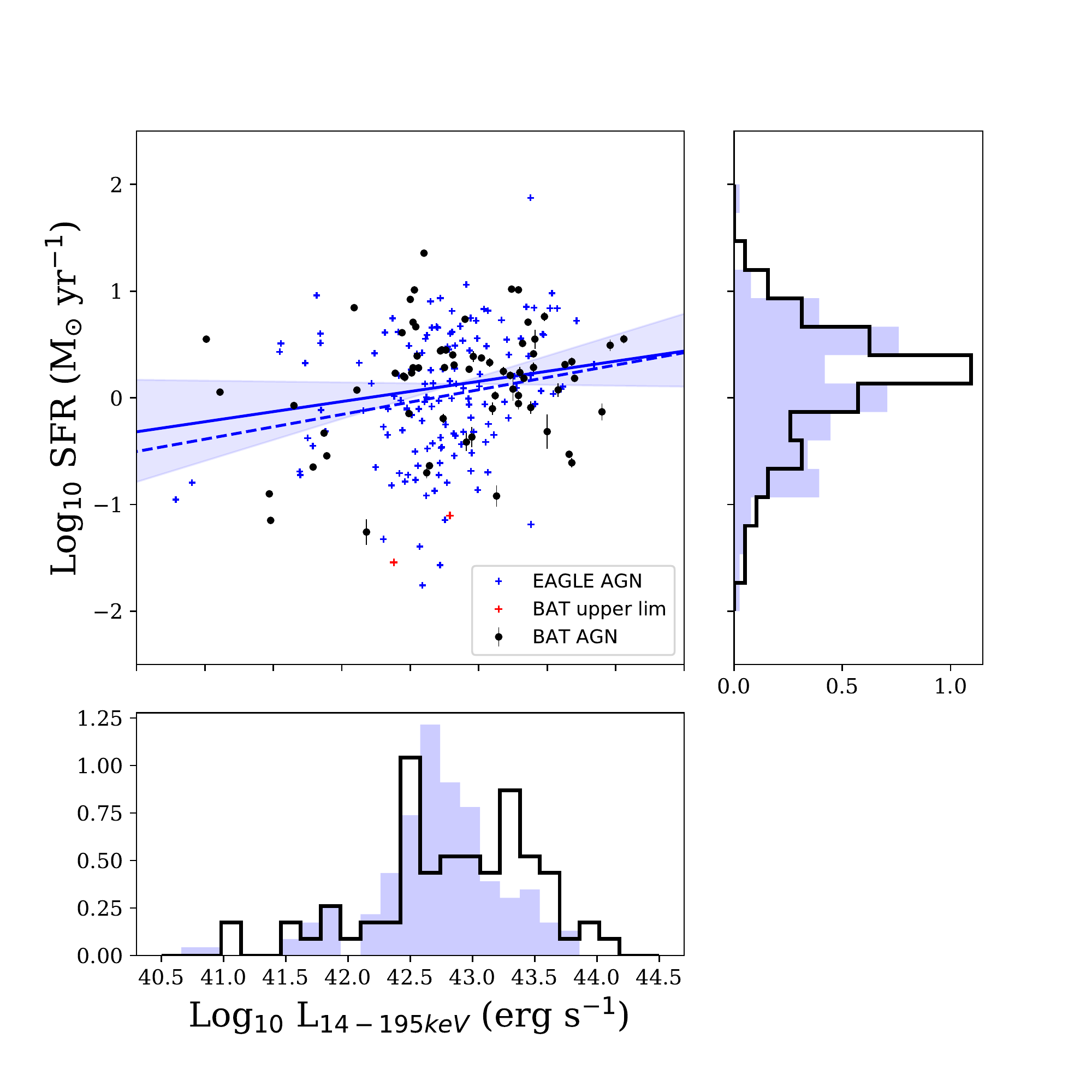}
	\caption{SFR versus X-ray luminosity for the BAT AGN (black circles) and a single realisation of the EAGLE AGN (blue crosses) samples. The solid blue line shows the best fitting line between SFR and X-ray luminosity recovered from {\sc Linmix} with 1$\sigma$ error intervals given by the shaded regions for the BAT AGN sample. The dashed line indicates the {\sc Linmix} best fitting for the single realisation of the EAGLE AGN sample. The sub panels show the BAT AGN SFR and X-ray luminosity distributions in black and the distributions of the single realisation of the EAGLE AGN sample in shaded blue. We see good agreement in the SFR and X-ray luminosity distributions and a good replication by EAGLE of the expected trend between SFR and X-ray luminosity.}
	\label{fig:SFR_v_Lx}
\end{figure}

In order to investigate how well EAGLE reproduces the star forming properties of local AGN galaxies we directly compare our BAT AGN and a single realisation of the EAGLE AGN sample. The first key tests are EAGLEs predictions of the trends between SFR and X-ray luminosity and the respective distributions: any major discrepancies indicates possible problems with the treatment of AGN in the simulation. The trends between these two variables can also be used to assess whether previously found trends are reproduced in both the observational and simulated data sets \citep{2012A&A...545A..45R, 2015MNRAS.453..591S, 2017MNRAS.466.3161S}.

Figure~\ref{fig:SFR_v_Lx} shows the results of X-ray luminosity versus SFR for a single realisation of our EAGLE sample. We find good overall agreement between the two samples, with no distinct outliers. 

In order to reveal trends between the SFR and X-ray luminosity for the BAT AGN sample, we use {\sc Linmix}, a Bayesian linear regression code originally written in IDL \citep{2007ApJ...665.1489K}. It has been shown to outperform other fitting codes, accounts for errors and also allows for upper limits and non-detections in the estimations. A {\sc python} version was created and is publicly available on GitHub\footnote{\url{https://github.com/jmeyers314/linmix}}.

\begin{figure*}
	\centering
	\includegraphics[width=1.8\columnwidth]{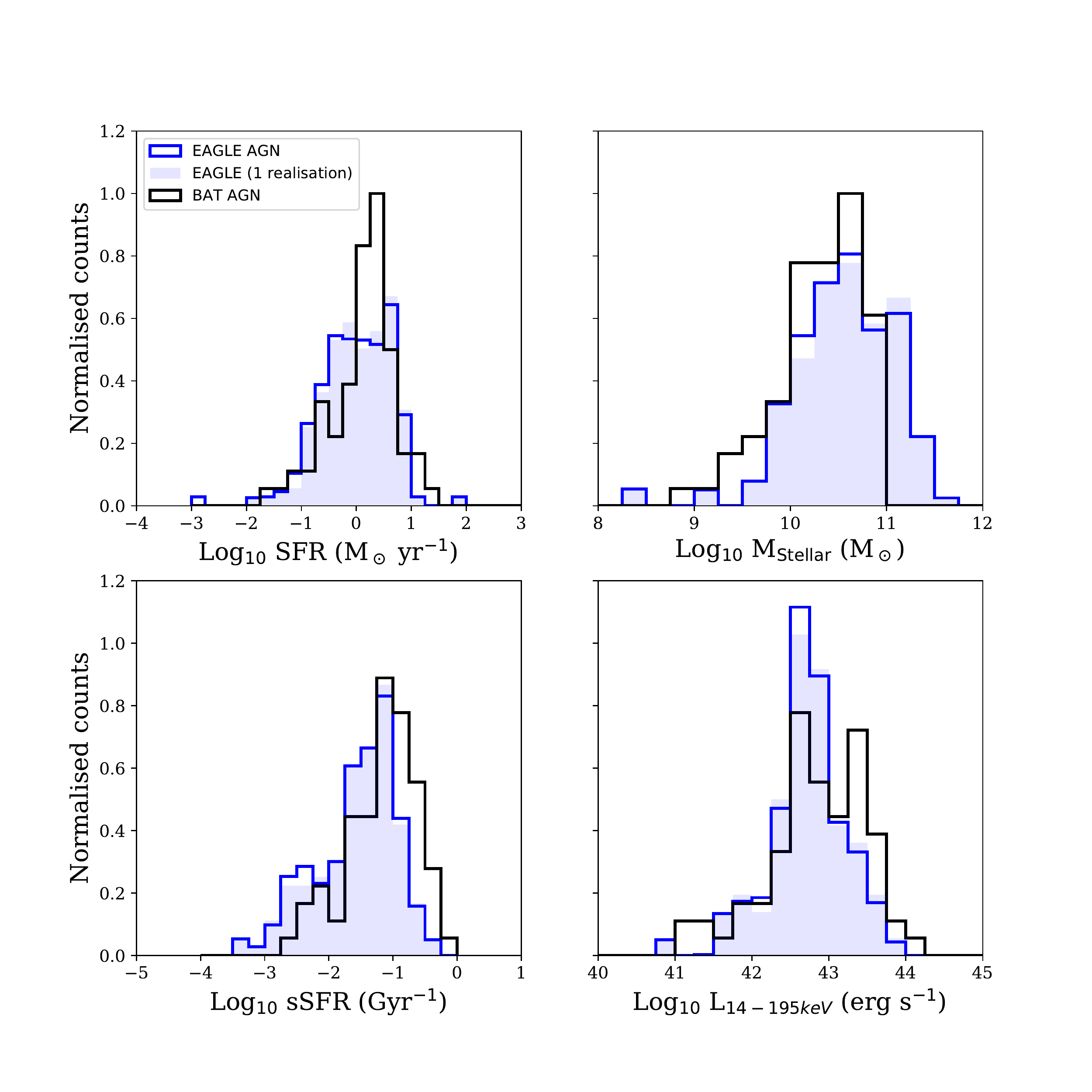}
	\caption{The distributions in SFR, stellar mass, sSFR and X-ray luminosity of our BAT AGN sample (black line), full EAGLE AGN sample (blue line, see Section~\ref{subsec:sim_sample} for details) and a single realisation of the EAGLE AGN (shaded blue, also Section~\ref{subsec:sim_sample}). We see good agreement in the SFRs and X-ray luminosities, however on average higher stellar masses in the full EAGLE AGN sample and therefore lower sSFRs.}
	\label{fig:ssfr_dist}
\end{figure*}

The trend yielded by the fitting process for this specific single realisation is represented in Figure~\ref{fig:SFR_v_Lx} by the solid blue line with 1$\sigma$ uncertainties given by the shaded areas. We find a slight trend of increasing SFR with X-ray Luminosity, as seen in \citet{2017MNRAS.466.3161S}. When we run {\sc Linmix} on our EAGLE AGN sample we find that EAGLE (dashed blue line) recovers the trend well and within the uncertainties (shaded regions). 

The median, maximum and minimum X-ray luminosities of the EAGLE AGN lie $\approx$ 0.1 dex lower than the BAT AGN, within the observational uncertainties. The SFR distributions (Figure~\ref{fig:SFR_v_Lx}, right hand panel) are in good agreement for this single choice of realisation, as can be seen in the right hand panel, with the differences in the median within 0.1 dex and with similar ranges. Most realisations of the EAGLE AGN yield similar results; we perform a further quantitative statistical analysis in Section~\ref{res: ssfr}. 

\begin{table*}
	\centering
	\caption{The median values of respective host galaxy properties (with the respective 16$^{\mathrm{th}}$ and 84$^{\mathrm{th}}$ percentiles in brackets) for the BAT AGN sample, a single realisation of the EAGLE AGN (EAGLE AGN (1)) and the EAGLE AGN sample (full 10,000 realisations). The probability yielded by a K-S test between 1 realisation and the BAT AGN are given in column 5 (with the supremum in brackets) and the full 10,000 realisations and the BAT AGN in column 6 (median, 16$^{\mathrm{th}}$ and 84$^{\mathrm{th}}$) for the probabilities and column 7 (median, 16$^{\mathrm{th}}$ and 84$^{\mathrm{th}}$) for the respective suprema values.}
	\label{tab:dists}
	\begin{tabular}{lcccccr}
		\hline
		Host Galaxy Property & BAT AGN & EAGLE AGN (1) & EAGLE AGN & K-S (1 realisation) & K-S Probability & K-S Suprema \\
		& Med (16$^{\mathrm{th}}$, 84$^{\mathrm{th}}$) & Med (16$^{\mathrm{th}}$, 84$^{\mathrm{th}}$) & Med (16$^{\mathrm{th}}$, 84$^{\mathrm{th}}$) & \%, (Supremum) & Med (16$^{\mathrm{th}}$, 84$^{\mathrm{th}}$) & (corresponding) \\
		\hline
		Log$_{10}$ SFR (M$_\odot$ yr$^{-1}$) & 0.2 (-0.5, 0.6) & 0.1(-0.7, 0.6) & 0.0 (-0.7, 0.6) & 10.4\% (0.18) & 6.5\% (4.4\%, 10.6\%) & 0.19 (0.20, 0.18) \\
		& & & & & & \\
		Log$_{10}$ M$_{\mathrm{Stellar}}$ (M$_\odot$) & 10.4 (9.8, 10.7) & 10.6 (10.1, 11.1) & 10.6 (10.1, 11.1) & 0.3\% (0.23) & 0.7\% (0.4\%, 1.1\%) & 0.24 (0.23, 0.25) \\
		& & & & & & \\
		Log$_{10}$ sSFR (Gyr$^{-1}$) & -1.0 (-1.7, -0.7) & -1.5 (-2.4, -1.0) & -1.5 (-2.4, -1.0) & 0.01\% (0.32) &  0.02\% (0.01\%, 0.03\%) & 0.30 (0.32, 0.29) \\
		& & & & & & \\
		Log$_{10}$ L$_{14-195 \mathrm{keV}}$ (erg s$^{-1}$) & 42.9 (42.4, 43.4) & 42.8 (42.4, 43.2) & 42.7 (42.3, 43.2) & 2.3\% (0.21) &  1.1\% (0.7\%, 1.7\%) & 0.23 (0.22, 0.24) \\
		\hline
	\end{tabular}
\end{table*}

\subsection{Comparison of host galaxy distributions}
\label{res: ssfr}

We then compare the overall distributions in SFR, stellar mass, sSFR and X-ray luminosity of the 10,000 realisations of the EAGLE AGN with the respective BAT AGN distributions, shown in Figure~\ref{fig:ssfr_dist}. The solid black line represents the host galaxy property distributions of the BAT AGN, the solid blue line gives the host galaxy distributions of the EAGLE AGN sample after 10,000 realisations and the shaded blue shows the distribution of a single realisation of the EAGLE AGN sample, in order to provide a comparison of the variance of a single realisation to the average. The median of each host galaxy property distribution for all three samples as well as the 16$^{\mathrm{th}}$ and 84$^{\mathrm{th}}$ percentiles are given in Table~\ref{tab:dists}. 

In order to statistically compare the similarity of the distributions of both a single realisation and the EAGLE AGN distribution to the BAT AGN we performed a Kolomogorov-Smirnov (K-S) test using the \textsc{SciPy} implementation. From the calculated supremum, a p-value is derived, providing a statistic of the probability that the two distributions are drawn from the same parent distribution. We choose a threshold of 5\% (a p-value of 0.05 or below) that the two distributions of host galaxy properties could be drawn from the same parent distribution. The results of this test for both one realisation and the median, 16$^{\mathrm{th}}$ and 84$^{\mathrm{th}}$ percentiles for the full 10,000 realisations, with the associated suprema values, can also be found in Table~\ref{tab:dists}.

Although the probabilities given by a K-S test for most distributions are relatively low, the probabilities in the SFR distributions is above the 5\% threshold we set. The median, 16$^{\mathrm{th}}$ and 84$^{\mathrm{th}}$ percentiles of the distributions in SFR and X-ray luminosity are in agreement, with differences of $\sim$ 0.1 in the medians and similar values in the percentiles. These values are within expected observational uncertainties. We see a difference in the medians of the stellar mass of 0.2 dex. This combined with the differences in the SFR estimations, causes $\sim$ 0.4 dex differences in the sSFR distributions. This leads to very low probability yielded by a K-S test that the two distributions are drawn from the same parent distribution.

By investigating the position of AGN host galaxies relative to the SF-MS, we may reveal clues about their evolutionary stage, and thereby how an AGN may be affecting a host galaxy. By also investigating the width of the sSFR distributions, we can also investigate possible signs of AGN feedback as described in \citet{2018MNRAS.475.1288S}, whereby implementing AGN feedback in the EAGLE simulations drives broader sSFR distributions in the galaxy population than when AGN feedback is not implemented.

\begin{figure}
	\centering
	\includegraphics[width=\columnwidth]{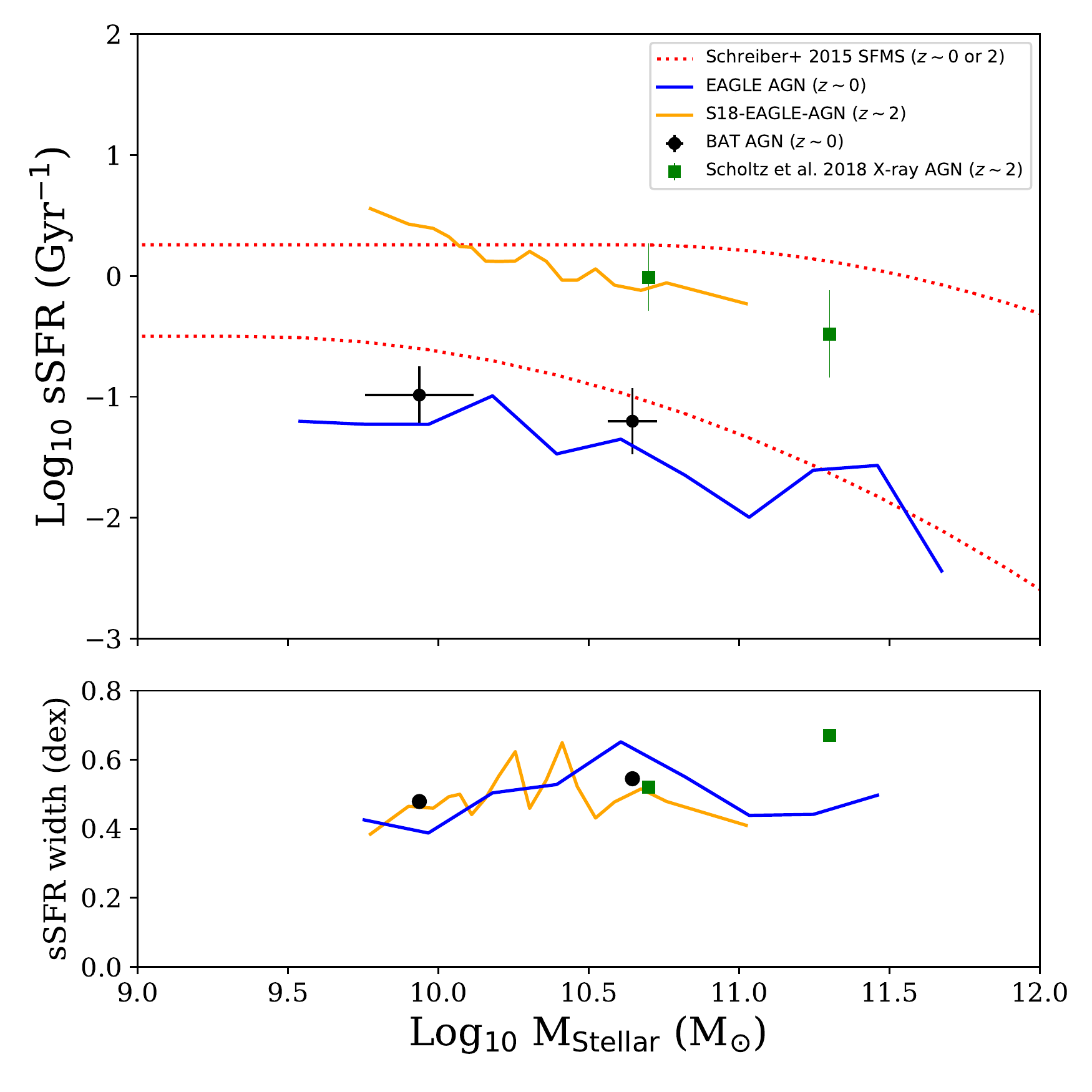}
	\caption{Top panel: sSFR versus stellar mass for the BAT AGN sample (black points), a single realisation of the EAGLE AGN (blue line), observed AGN from \protect\citet{2018MNRAS.475.1288S} (green points), simulated AGN from \protect\citet{2018MNRAS.475.1288S} (orange line) and the respective SF-MS (red dotted lines). The median of the BAT AGN lie below the SF-MS, in agreement with the findings from \protect\citet{2018MNRAS.475.1288S}. The single realisation of the EAGLE AGN also lies below the SF-MS, however lower than expected. This indicates that EAGLE underpredicts the expected sSFRs of our BAT AGN. Bottom panel: The width of the sSFR distributions for the samples described above.}
	\label{fig:ssfr_jan}
\end{figure}

Following \citet{2018MNRAS.475.1288S}, we split the BAT AGN sample into low and high stellar mass bins using the median stellar mass and then took the median values of the sSFR distributions. For the EAGLE AGN sample, we split the samples into 16 stellar mass bins, each of 0.2 dex ranging in Log$_{10}$ M$_{\mathrm{Stellar}}$ $=$ 9 - 12 M$_{\odot}$ and plotted the median. These values were then compared to the SF-MS according to the prescription of \citet{2015A&A...575A..74S}. We also calculate the width of the sSFR distribution of each bin, excluding those bins with only 1 galaxy. These results can be seen Figure~\ref{fig:ssfr_jan}. We also compare this to the results found in figure 6 of \citet{2018MNRAS.475.1288S}, to see if the two studies yield similar results despite utilising differently selected samples of AGN at different redshifts.

In the top panel, the BAT AGN samples (black points) lie below the $z \sim 0$ SF-MS (lower red dotted line) by 0.4 and 0.3 dex respectively at constant stellar mass, outside of observational uncertainties and the expected scatter of $\sim$ 0.3 dex in the SF-MS \citep{2015A&A...575A..74S}. The EAGLE AGN sample (blue line) lies $\approx$ 0.5 dex below the SF-MS at most stellar masses with a difference of 0.6 dex and 0.4 dex lower at the same stellar masses as the two BAT AGN points. We compare this to \citet{2018MNRAS.475.1288S}, where the observed AGN (hereafter S18 X-ray AGN, green squares) lie 0.3 dex and 0.6 dex below the $z \sim 2$ main sequence (upper red dotted line). The S18-EAGLE-AGN (orange line) are 0.4 dex below at a stellar mass of 10$^{10.6}$ M$_{\odot}$, the median stellar mass of the low mass S18 X-ray AGN sample. This shows the behaviour in the observations and simulations is similar across different reshifts and selections, however that the differences in the sSFR distributions found in Section~\ref{res: ssfr} translate to the sSFR vs mass trends, whereby the EAGLE AGN show lower sSFRs than expected for the observations.

\begin{table*}
	\centering
	\caption{The median, 16$^{\mathrm{th}}$ and 84$^{\mathrm{th}}$ percentiles of each distribution from our MC analyses for each host galaxy property in our BAT AGN sample, a single run of the MC analyses and the full 10,000 iterations of the MC analyses. The last three columns show the associated K-S test statistics between the BAT AGN distributions and one iteration of the MC analyses and the BAT AGN distributions and the distribution from 10,000 iterations of the MC analyses.}
	\label{tab:MC_dists}
	\begin{tabular}{lcccccr}
		\hline
		Host Galaxy Property & BAT AGN & EAGLE AGN (1) & EAGLE AGN & K-S (1 realisation) & K-S Probability & K-S Suprema \\
		& Med (16$^{\mathrm{th}}$, 84$^{\mathrm{th}}$) & Med (16$^{\mathrm{th}}$, 84$^{\mathrm{th}}$) & Med (16$^{\mathrm{th}}$, 84$^{\mathrm{th}}$) & \%, (Supremum) & Med (16$^{\mathrm{th}}$, 84$^{\mathrm{th}}$) & (corresponding) \\
		\hline
		Log$_{10}$ SFR (M$_\odot$ yr$^{-1}$) & 0.2 (-0.5, 0.6) & 0.1(-0.5, 0.7) & 0.1 (-0.7, 0.7) & 45.1\% (0.12) & 43.2\% (24.1\%, 62.4\%) & 0.13 (0.15, 0.11) \\
		& & & & & & \\
		Log$_{10}$ M$_{\mathrm{Stellar}}$ (M$_\odot$) & 10.4 (9.8, 10.7) & 10.5 (10.1, 11.1) & 10.6 (10.1, 11.1) & 1.3\% (0.23) & 0.3\% (0.1\%, 1.8\%) & 0.27 (0.29, 0.24) \\
		& & & & & & \\
		Log$_{10}$ sSFR (Gyr$^{-1}$) & -1.0 (-1.7, -0.7) & -1.4 (-2.2, -1.0) & -1.4 (-2.3, -1.0) & 0.1\% (0.28) &  0.08\% (0.02\%, 0.2\%) & 0.29 (0.31, 0.27) \\
		& & & & & & \\
		Log$_{10}$ L$_{14-195 \mathrm{keV}}$ (erg s$^{-1}$) & 42.9 (42.4, 43.4) & 42.8 (42.5, 43.3) & 42.8 (42.5, 43.4) & 5.9\% (0.19) &  10.1\% (5.9\%, 14.9\%) & 0.18 (0.19, 0.17) \\
		\hline
	\end{tabular}
\end{table*}

In the bottom panel we see the width of the sSFR distribution as a function of stellar mass. We see similar results between the simulations at different redshifts, which both display similar widths ranging between $\sim$ 0.4 dex and 0.6 dex at all stellar masses, with no obvious trend with stellar mass. We see that the widths from the observational points are within 0.1 dex of the simulated data, in good agreement. 

\subsection{Monte-Carlo methods}
\label{res: MC}

Many modestly luminous simulated AGN in EAGLE are excluded from our analysis by our application of the BAT survey limits calculated from the central point of the simulation box (see Section~\ref{subsec:sim_sample} for details). Due to the relatively small size of the simulation box (100 cMpc per side), this could introduce biases due to cosmic variance. In order to check that the centre of the simulation box (or indeed any point in the simulation box) is not biased, we applied Monte-Carlo (MC) methods to draw multiple samples of AGN from the EAGLE box.

By assuming the distribution of galaxies in the total EAGLE volume is approximately homogeneous and isotropic we can model the number of galaxies per bin in proper distance as a power law. This is not unreasonable as we expect only one massive cluster within the observational volume we are using and therefore that our sample is dominated by field AGN and galaxies. To do this, we use the random number generator routine from \textsc{NumPy} based on a power law to generate a random distribution of distances with the amount of galaxies in the simulation box above a stellar mass of 10$^9$ M$_\odot$ and a maximum distance matched to our BAT AGN sample, 67.8 Mpc. We then assign one of these randomised distances to each galaxy. Once the distances are assigned, assuming the same Planck cosmology applied to EAGLE, we can calculate the expected AGN X-ray flux from each galaxy. This allows us to repeat the same flux correction due to X-ray obscuration described in Section~\ref{subsec:sim_sample} on the new X-ray flux values and select a new sample of AGN from the simulation as if observing from a different area of the simulation box, while saving significantly on computational power. 

This technique allows us to randomly draw a different sample of lower luminosity AGN while retaining a large number of higher luminosity AGN which we would expect to see from most positions in the simulation box, thereby utilising the full statistical power of EAGLE. We then compare the medians and scatter of the distributions and run a K-S test on the BAT AGN SFR (sSFR, stellar mass, X-ray luminosity) distribution and the respective newly drawn EAGLE distributions in order to recover the probability of the two distributions being drawn from the same parent sample. We run this process of random assignment, calculation, application of the BAT survey limits and comparison 10,000 times, collecting the host galaxy property distributions and the probabilities (and associated suprema values) between the observational sample and each run of the MC analyis. The median of each host galaxy property distribution, with the 16$^{\mathrm{th}}$ and 84$^{\mathrm{th}}$ percentiles given in the first 3 columns of Table~\ref{tab:MC_dists}. We also show the K-S results from a comparison of the single realisation of the EAGLE AGN via MC analyses and the median probabilities with 16$^{\mathrm{th}}$ and 84$^{\mathrm{th}}$ percentiles from the full 10,000 realisations.

Table~\ref{tab:MC_dists} shows that differences in the median and scatter of the SFR and X-ray luminosity distributions of the two samples are unchanged and remain of the order of 0.1 dex, within the observational uncertainties. The median probabilities of the two samples being drawn from the same parent distribution are also higher at 45\% and 10\% respectively. The differences in the stellar mass distributions and therefore the sSFR distributions remain, however, with differences of $\sim$ 0.2 dex in the stellar mass distributions and 0.4 dex in the sSFR distributions, with low percentages of agreement from the K-S tests. These results show that comparing the BAT AGN to the EAGLE AGN in the centre of the box or drawing random samples is relatively unbiased as far as the SFRs, sSFRs and stellar masses are concerned. The X-ray luminosities, however, go from below our level of acceptance provided by a K-S test to above the threshold; we discuss this further in Section~\ref{subsec:disc host}.

\subsection{The evolution of high and low-z AGN in EAGLE}
\label{res: trees}

\begin{figure*}
	\centering
	\includegraphics[width=1.6\columnwidth]{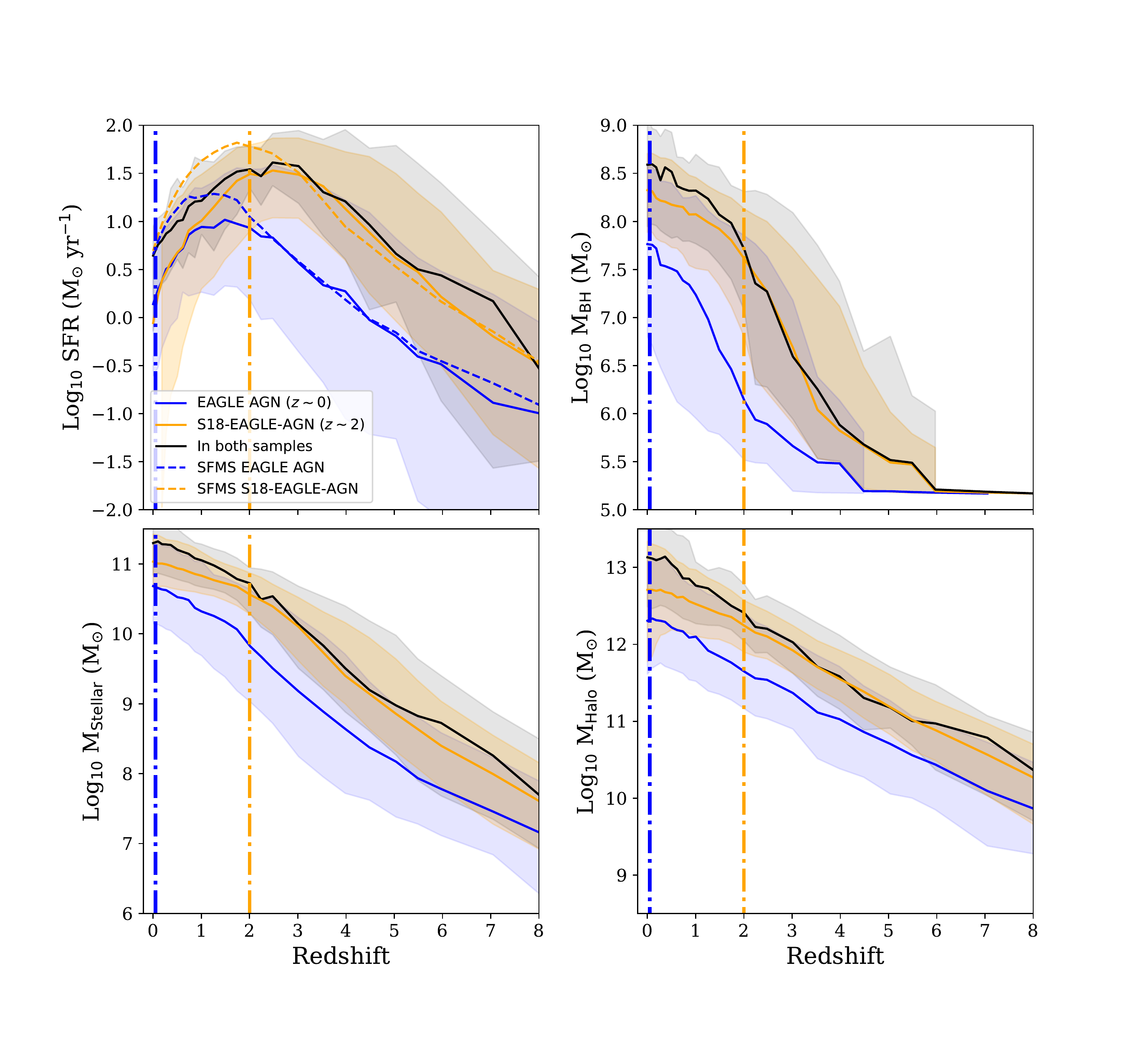}
	\caption{The evolution from $z = 8$ to $z = 0$ of the AGN comparison sample used in this project (blue) and in \protect\citet{2018MNRAS.475.1288S} (orange) with those host galaxies contained in both samples in black. Median values of the population are given by solid lines and 16$^{\mathrm{th}}$ and 84$^{\mathrm{th}}$ percentiles by the shaded regions. The top left panel shows the average SFR of each sample with the dashed lines representing the respective SF-MS for each sample. The top right panel shows the evolution of the black hole mass (seeded at 10$^{5}$ M$_{\odot}$, causing a flat evolution at higher redshifts). The bottom panels show the evolution of the stellar mass (left) and the halo mass (right). The dot-dashed vertical lines show the epoch of selection of the respective samples.}
	\label{fig:traceback}
\end{figure*}

Figure~\ref{fig:traceback} shows the results of our investigation into the evolution of various characteristic host galaxy properties between $z = 0$ and $z = 8$ of our EAGLE AGN sample and the S18-EAGLE-AGN sample as well as those AGN host galaxies contained in both samples; i.e., those systems that display significant AGN activity above the X-ray flux limit at both $z = 0$ and high redshift. In the top left hand panel we can see that our EAGLE AGN sample (blue solid line) follows the respective SF-MS (blue dashed line) very tightly until $z \simeq 2.5$. We calculate the value of the SF-MS at all points using the redshift and median stellar mass, according to the prescription in \citet{2015A&A...575A..74S}. At this point the average SFR decreases slowly relative to the main sequence until we reach the present day. Our average sample lies significantly (0.5 dex) below the SF-MS at $z = 0$. Similarly with the S18-EAGLE-AGN (orange solid line), the galaxies are on the respective SF-MS (orange dashed line) until significant quenching starts to have an effect at $z \simeq 3.5$, by $z = 0$ the average offset of the S18-EAGLE-AGN from the SF-MS is $\sim$ 0.6 dex.  

Having established that the star formation is indeed quenching at different epochs for the two simulated AGN samples, we now try to establish the host galaxy properties influencing or following this behaviour. In the top right hand panel we have the black hole mass as a function of redshift. We see that black holes are seeded at 10$^{5}$ M$_{\odot}$ and undergo very little initial growth. The black holes then reach a point at which they grow very rapidly and non-linearly. This phase is followed by a weakening of this rate of growth towards low redshifts. This behaviour is in agreement with previous findings in \citet{2017MNRAS.465...32B} and \citet{2018MNRAS.481.3118M}, whereby they propose a three stage growth scenario of black holes within EAGLE which is halo mass and redshift dependent. First, an initial stellar feedback dominated phase, where the black hole experiences little growth due to a mixture of stellar and supernovae feedback inhibiting gas accretion onto the black hole. Second, a non-linear rapid black hole growth phase, where the gravitational potential well of the galaxy overcomes the stellar and supernovae feedback and causes significant amounts of gas to migrate to the centre of the galaxy, and finally an AGN feedback regulated growth phase where the rate of growth slows.

In the bottom panels of Figure~\ref{fig:traceback} we see the growth of the stellar (left hand panel) and halo mass (right hand panel). The stellar mass grows non-linearly, with a slight correlation with the black hole growth, i.e. the change in growth gradient occurs at the same points as for the black hole, although this behaviour is much weaker. The rate of growth is significantly less at low redshifts, especially for the S18-EAGLE-AGN host galaxies. The halo masses grow fairly linearly throughout redshift, however there are hints of a deceleration of this growth at very low redshifts. 

The black line represents those galaxies contained in both samples as described in Section~\ref{subsec:SFH}, i.e. galaxies which display significant AGN activity at both high and low redshift within the simulation (23\% of all AGN galaxies selected at $z = 0$). We notice that their behaviour is similar to that of the S18-EAGLE-AGN sample at high redshifts. At low redshifts, however, these objects appear to continue growing in black hole mass, stellar mass and halo mass. The median of the SFR of this sample is 0.7 dex above the median value of the S18-EAGLE-AGN, with a 0.3 dex difference in the black hole mass, 0.3 dex difference in the stellar mass and a 0.4 dex difference in the halo mass. With an average black hole mass of 10$^{8.6}$ M$_{\odot}$, stellar mass of 10$^{11.3}$ M$_{\odot}$ and halo mass of 10$^{13.1}$ M$_{\odot}$, these objects are some of the most massive in the simulation. These differences and potential drivers are discussed further in the following section.

\section{Discussion}
\label{sec:disc}

In this research we have taken a sample of observed AGN (the BAT AGN sample) and compared them to a volume and flux matched sample of simulated AGN (the EAGLE AGN sample). We have shown that the level of agreement between the SFR and X-ray luminosity distributions of our BAT AGN and EAGLE AGN is reasonable, both from our analysis utilising the centre of the simulation box as an observer view point and our methods utilising MC methods. The sSFR distributions are less well reproduced, driven by differences in the stellar masses of the host galaxies, whereby EAGLE overpredicts, on average, the host galaxy stellar mass by 0.2 dex. Both observational and simulated samples lie below the SF-MS, in agreement with previous studies \citep{2018MNRAS.475.1288S}, with similar widths in the sSFR distributions. We also explored the characteristic evolution of host galaxy properties of AGN selected at different redshifts through the use of merger trees in EAGLE. AGN host galaxies display similar evolutionary behaviour over different redshifts, whereby downsizing is present. We also find that a number of AGN selected at high redshift are also present in the sample selected at low redshift within the simulations. We will now put these results in to context and discuss their meaning. We will first discuss the comparison of the simulated and observed host galaxy properties including possible reasons for any disagreements between the two in Section~\ref{subsec:disc host}. We then discuss the conversion factor used to convert the BHAR from EAGLE into simulated X-ray luminosities and how this may affect number counts and our results in Section~\ref{subsec:disc counts}. Finally we discuss in more detail the evolution of AGN host galaxies within EAGLE and the drivers behind this evolution in Section~\ref{subsec:comp scholtz} 

\subsection{Comparison of BAT and EAGLE AGN host galaxy properties}
\label{subsec:disc host}

We find reasonable agreement between the SFR distributions of our volume and flux matched BAT AGN and EAGLE AGN comparison samples, with the caveat that we applied an offset of 0.2 dex to the SFRs, following previous studies \citep[e.g.][]{2016A&C....15...72M, 2018MNRAS.481.3118M, 2018MNRAS.475.1288S}. The median and 16$^{\mathrm{th}}$ and 84$^{\mathrm{th}}$ percentiles of the SFR distributions are similar and the statistics recovered from a K-S test provide a moderate probability that the two SFR distributions are drawn from the same parent sample.
	
We find $\sim$ 0.1 dex differences in the median and scatter of the X-ray luminosity distributions, and a weak probability that the two distributions are drawn from the same sample, based on a K-S test. We note that although the probability yielded by a K-S test is low, the differences of 0.1 dex in the median and scatter of the distributions are within the observational uncertainties, which should be the main indicator of the similarity of the distributions.
	
The sSFR distributions are less well reproduced (differences of $\sim$ 0.4 dex in the median and scatter of the distributions). This is caused by EAGLE, on average, overpredicting the stellar masses of AGN host galaxies by $\gtrsim$ 0.2 dex, as seen from the distributions in Figure~\ref{fig:ssfr_dist}. We note, however, that observational uncertainties of the stellar masses are of the order of 0.2 dex, which could contribute to some of the differences between the two distributions. K-S tests yield very low probabilities that the observed and simulated distributions are drawn from the same parent sample for both the stellar mass and sSFR. We also note that these differences would be increased had we not applied the 0.2 dex offset in the SFR.

The results from this comparison of the AGN host galaxy property distributions are reinforced by our MC methods, which were carried out to verify that the central point of observation in the simulation box is not biased. The differences in the median and scatter of the X-ray luminosity and SFR distributions remain at $\sim$ 0.1 dex and within the observation uncertainties. The probability that the SFRs are drawn from the same parent sample are higher. The differences in the median and scatter of the sSFR and stellar mass distributions remain when we consider the results from the MC methods compared to selecting our sample from the centre of the simulation box. The probabilities of the sSFR and stellar mass distributions being drawn from the same parent sample remain low, justifying our conclusion that these results hold regardless of the point of observation in the simulation box.

The X-ray luminosity distributions, however, go from a low probability of being drawn from the sample parent sample in the centre of the simulation box ($\sim 1$ per cent) to a higher one from our MC methods ($\sim 10$ per cent, which is above our 5 per cent threshold). To test if the centre of the simulation box is biased in the X-ray luminosities we selected AGN using the same flux matching techniques as earlier, however from the eight corners of the simulation box, with a distance cut of 70 Mpc. This re-creates the {\it Swift}-BAT volume, but from a different set of independent vantage points in the simulation and thereby check if the the centre of the simulation box is biased. When we run the same techniques and tests, we find that the probabilities of the SFR, sSFR and stellar mass distributions being drawn from the same parent sample remain similar to before, however the probabilities for the luminosity distributions are 5.2 per cent. We postulate that the change in probabilities could be due to two scenarios. First that there is a slight bias in the centre of the simulation box compared to other locations within the simulation box. Second that the Milky Way may be in a biased point within the nearby universe compared to the simulation box. A thorough investigation, however, is beyond the scope of this study.
	
One scenario which could explain the 0.2 - 0.4 dex difference in the average stellar masses (and thereby tensions of 0.4 dex in the average sSFRs) found in our study has been postulated in a recent study carried out by \citet{2020ApJ...893..111L}. They compare stellar mass estimates using parametric SFHs to those estimated using non-parametric SFHS. They find that parametric SFHs may underestimate empirically derived stellar masses in observations on the order of 0.1 - 1 dex, which could account for the $\gtrsim$ 0.2 dex differences in median and scatter of the observed and simulated stellar masses distributions. This result also affects the observed stellar mass function, so future work in this field is also likely to have implications to the modelling of simulations, namely that if stellar masses are systematically underestimated then the stellar mass function needs revising. As this is one of the most commonly used calibrators for simulations, this would require changes in the sub-grid physics.

An alternative solution to the tensions in the stellar masses can be provided by \citet{2018MNRAS.481.3118M}. This study found that the rapid growth phase of SMBHs in EAGLE was triggered by increasing fractions of major mergers towards low redshifts. Major mergers can deliver significant amounts of stellar mass to the central galaxy, thereby biasing the comparison sample towards higher stellar masses. If there are indeed differences in the fueling mechanisms between the observations and simulations (such as if AGN fuelling is triggered less frequently by mergers in the observed universe), this could be a source of the biases in the stellar masses in simulated AGN host galaxies. An in depth comparison of AGN fueling mechanisms between simulations and observations, however, is beyond the scope of this work.

\subsection{Effects of the bolometric correction and number counts of AGN}
\label{subsec:disc counts}    

As stated in Section~\ref{subsec:sim_sample}, once we apply our X-ray luminosity correction, we find that EAGLE predicts 132 AGN per sample on average over 10,000 realisations, compared to the 72 in our BAT AGN sample, an overestimation in the number counts. We note that there are assumptions we have made that may impact these results. Our conversion of the fraction of luminosity from an AGN (L$_{\mathrm{bol}}$) radiated in the Ultra hard X-ray band is taken from \citet{2017MNRAS.470..800T} (a value of 8.5). When we assume a bolometric conversion such as that applied by \citet{2017ApJ...850...74K} (with a value of 8), we find an average of 142 AGN over 10,000 realisations. Conversely, when we assume a bolometric correction factor of 10 such as that in \citet{2018MNRAS.475.1288S}, albeit for a different X-ray energy band, an average of 120 AGN is yielded over 10,000 realisations. This shows that the AGN counts are extremely sensitive to the bolometric to ultra hard X-ray conversion factor used. 

We also note that the X-ray luminosity correction from \citet{2015ApJ...815L..13R} accounts for fractions of AGN at various opening angles, which is not accounted for in the simulations. Finally, EAGLE itself could be overpredicting the number count of AGN, similar to overpredicting the stellar mass of the host galaxy. The source of the differences would need to be investigated in further work.

Although the different bolometric conversions yield varying counts of AGN on average in the sample selection, when implemented and the simulated samples compared to the BAT AGN sample, we find very little difference in the results. The difference in the percentages yielded by a K-S test for the probability of the two samples being drawn from the same parent distribution change by less than 3\% in the SFR distributions, 0.5\% for the X-ray luminosity, 0.1\% for the sSFR distributions and 0.3\% for the stellar mass, in both the original method and in the MC methods. Similarly the medians of SFR, sSFR and stellar mass distributions change by $<$ 0.1 dex, not significantly different to the results from our original comparison.

\subsection{The difference in host Galaxies of AGN at different epochs}
\label{subsec:comp scholtz}

From our use of the merger trees for AGN samples selected at different epochs, our EAGLE AGN sample and the S18-EAGLE-AGN sample, we see how the selection criteria can affect the expected host galaxy properties of AGN and how similarly their host galaxy properties can evolve. Downsizing, whereby less massive objects undergo physical processes later in the history of the universe \citep{2005ARA&A..43..827B}, is clearly displayed in the SFRs, black hole masses, stellar masses and halo masses, as shown in Figure~\ref{fig:traceback}. 

The EAGLE AGN (which have on average smaller stellar or halo masses across all epochs) leave the SF-MS later than the S18-EAGLE-AGN ($z = 3.5$ compared to $z = 2.5$), in other words their star formation is quenched at later times. The black hole masses in the EAGLE AGN sample also grow at later epochs than the S18-EAGLE-AGN, especially their phase of most rapid growth. The point at which they start this rapid growth phase also approximately coincides with the point at which both populations leave the SF-MS. This shows the connection between the time at which the AGN injects proportionally the most energy and the quenching of star formation, as expected. We also see this growth slow down at later epochs, confirming the overall behaviour of black hole growth seen by \citet{2018MNRAS.481.3118M}.  

We also see that the star formation starts deviating from the main sequence for both our EAGLE AGN sample and the S18-EAGLE-AGN when the halo masses reach a critical point of M$_{\mathrm{Halo}}$ $\approx$ 10$^{12}$ M$_\odot$. Studies by \citet{2017MNRAS.465...32B} and \citet{2017MNRAS.468.3395M} explain that the halo is massive enough that the stellar and supernovae feedback is no longer efficient enough to eject gas from the galaxy. At this point gas starts flowing into the centre of the galaxy, accreting around the black hole, thereby growing it. The black hole at this point then starts injecting energy back into the host galaxy, thereby quenching star formation. 

AGN galaxies which are contained in both samples and therefore display significant AGN activity at both high and low redshift are of interest, as this phenomena cannot be observed. We find that 23\% of all AGN galaxies selected at $z = 0$ displayed significant activity at high redshift. These objects tend to follow similar behaviour to the S18-EAGLE-AGN sample, however at low redshift they continue to grow in stellar, halo and black hole mass.  

One problem with some hydrodynamic simulations is the inability to fully reproduce the quenched fraction of massive galaxies, driven by higher gas fractions than in the observations \citep[see e.g.][]{2015A&C....13...12N}. A possible scenario this may lead to is a sub-sample of massive galaxies at high redshift able to undergo an AGN fuelling phase and then a low redshift continuing to be fuelled by gas, leading to higher than expected levels of star formation and accretion onto the SMBH. This then leads to higher BH, stellar and halo masses and an AGN phase at $z \sim 0$. We cannot fully rule out, however, that an AGN in the observed universe would display this behaviour due to the duty cycle or triggering from mergers.

\section{Conclusion}
\label{sec:Conc}

This work has compared the star forming properties of a sample of low redshift, hard X-ray selected AGN from the {\it Swift}-BAT all-sky survey. We used optical and IR photometry in conjunction with SED fitting in order to estimate host galaxy properties. We then compared the SFR, stellar mass, sSFR and X-ray luminosity distributions and position with respect to the SF-MS of this observational sample to a volume matched, flux matched sample of AGN from the EAGLE hydrodynamical suite of simulations.

We find a reasonable level of agreement in the SFR and X-ray luminosity distributions, with a $\sim$ 0.1 dex difference in the medians and scatter of the distributions, within the observational uncertainties. A K-S test recovers a median 6.5\% and 1.1\% probability that the observed and simulated distributions are drawn from the same parent sample respectively as observed from the centre of the simulation box. This improves 43.2\% and 10.1\% using MC methods to randomly draw AGN from the EAGLE simulation, meaning that the results are not significantly biased by the observers position in the simulation box. EAGLE appears to overpredict the stellar masses of the AGN comparison sample by 0.2 dex, which contributes to differences of 0.4 dex in the sSFRs, with low levels of probability that the observed and simulated distributions are drawn from the same parent sample as yielded by K-S tests. We find that the median of the sSFR distribution lies $\sim 0.3$ dex below the SF-MS and that the width of the distribution is $\sim0.4-0.6$ dex for both the observations and simulations, consistent with previous findings in \citet{2018MNRAS.475.1288S}.

We then compared the cosmic evolution of the host galaxy properties of our simulated EAGLE AGN with the simulated high redshift AGN sample from EAGLE in the work \citet{2018MNRAS.475.1288S}, in order to investigate if our results are similar with high redshift AGN and if AGN host galaxies selected at different epochs evolve in a similar manner. We find similar behaviour in the quenching of star formation in the simulations as previous studies, whereby the host galaxy star formation rate drops below the SF-MS at the time that the black hole goes through its phase of most rapid growth and that AGN selected at different cosmic epochs display similar quenching behaviour albeit at different times. We also find that a surprisingly high fraction of AGN observed at low redshift (23\%) underwent a significant unrelated accretion event at high redshift, as they are also selected as AGN at high redshift.

\section*{Acknowledgements}

The authors would like to thank the anonymous referee whose thorough reading and constructive comments have helped improve this study and manuscript.

TMJ would like to acknowledge the Pemberton Fund from the Durham Castle Society and the Leather Sellers guild for help in funding of a masters degree and this research at Durham University. 

DJR, DMA and RGB acknowledge the Science and Technology Facilities Council (grant codes
ST/P000541/1 and ST/T000244/1).

We thank the Virgo consortium and EAGLE team for making the data from their simulations public. 

This work used the DiRAC Data Centric system at Durham University,
operated by the Institute for Computational Cosmology on behalf of the
STFC DiRAC HPC Facility. This equipment was funded
by BIS National E-infrastructure capital grant ST/K00042X/1, STFC capital
grant ST/H008519/1, and STFC DiRAC Operations grant ST/K003267/1 and
Durham University. DiRAC is part of the National E-Infrastructure.

Funding for the Sloan Digital Sky Survey IV has been provided by the Alfred P. Sloan Foundation, the U.S. Department of Energy Office of Science, and the Participating Institutions. SDSS acknowledges support and resources from the Center for High-Performance Computing at the University of Utah. The SDSS web site is www.sdss.org.

SDSS is managed by the Astrophysical Research Consortium for the Participating Institutions of the SDSS Collaboration including the Brazilian Participation Group, the Carnegie Institution for Science, Carnegie Mellon University, the Chilean Participation Group, the French Participation Group, Harvard-Smithsonian Center for Astrophysics, Instituto de Astrofísica de Canarias, The Johns Hopkins University, Kavli Institute for the Physics and Mathematics of the Universe (IPMU) / University of Tokyo, the Korean Participation Group, Lawrence Berkeley National Laboratory, Leibniz Institut f\"{u}r Astrophysik Potsdam (AIP), Max-Planck-Institut f\"{u}r Astronomie (MPIA Heidelberg), Max-Planck-Institut f\"{u}r Astrophysik (MPA Garching), Max-Planck-Institut f\"{u}r Extraterrestrische Physik (MPE), National Astronomical Observatories of China, New Mexico State University, New York University, University of Notre Dame, Observatório Nacional / MCTI, The Ohio State University, Pennsylvania State University, Shanghai Astronomical Observatory, United Kingdom Participation Group, Universidad Nacional Aut\'{o}noma de M\'{e}xico, University of Arizona, University of Colorado Boulder, University of Oxford, University of Portsmouth, University of Utah, University of Virginia, University of Washington, University of Wisconsin, Vanderbilt University, and Yale University.

This project used data from 2MASS survey, a collaboration between The University of Massachusetts and the Infrared Processing and Analysis Center (JPL/ Caltech). Funding is provided primarily by NASA and the NSF.

This publication makes use of data products from the Wide-field Infrared Survey Explorer, which is a joint project of the University of California, Los Angeles, and the Jet Propulsion Laboratory/California Institute of Technology, funded by the National Aeronautics and Space Administration.

This project used data from the HyperLeda database (http://leda.univ-lyon1.fr).

This project used data from the NASA Sloan Atlas database. Funding for the NASA-Sloan Atlas has been provided by the NASA Astrophysics Data Analysis Program (08-ADP08-0072) and the NSF (AST-1211644).

This project used data from Pan-STARRs. The Pan-STARRS1 Surveys (PS1) and the PS1 public science archive have been made possible through contributions by the Institute for Astronomy, the University of Hawaii, the Pan-STARRS Project Office, the Max-Planck Society and all other associated institutions found on the database website. 

This project used data from the NASA Extragalactic database. Funded by NASA and other associated institutes.

\section{Data availability}

The data underlying this article are available from the public sources in the links or references given in the article (or references therein). Estimations from the article are given in the appendix.




\bibliographystyle{mnras}
\bibliography{AGN_bibliography} 



\appendix

\section{The BAT AGN sample}
\label{app: BAT AGN}

\begin{table*}
	\centering
	\caption{The BAT AGN Sample.}
	\label{tab:BAT AGN}
	\begin{tabular}{ l|c|c|c|c|c|r }
		\hline
		Swift-BAT ID & RA & Dec & Redshift & Log$_{10}$ L$_{14 - 195 \mathrm{keV}}$ (erg s$^{-1}$) & Log$_{10}$ SFR (M$_\odot$) & Log$_{10}$ M$_{\mathrm{Stellar}}$ (M$_\odot$ yr$^{-1}$) \\
		\hline
		SWIFTJ0048.8+3155 & 12.196 & 31.957 & 0.0149 & 43.90 & -0.13 $\pm$ 0.08 & < 10.15 \\
		SWIFTJ0059.4+3150 & 14.972 & 31.827 & 0.0149 & 43.13 & -0.92 $\pm$ 0.10 & 9.74 $\pm$ 0.14 \\
		SWIFTJ0111.4-3808 & 17.865 & -38.083 & 0.0118 & 43.38 & -0.092 $\pm$ 0.06 & < 9.84 \\
		SWIFTJ0114.5-3236 & 18.529 & -32.651 & 0.0120 & 42.72 & 0.44 $\pm$ 0.03 & 10.41 $\pm$ 0.08 \\
		SWIFTJ0201.0-0648 & 30.277 & -6.815 & 0.0136 & 43.66 & -0.53 $\pm$ 0.02 & 10.80 $\pm$ 0.04 \\
		SWIFTJ0241.3-0816 & 40.270 & -8.256 & 0.005 & 42.18 & -1.26 $\pm$ 0.12 & 10.48 $\pm$ 0.08 \\
		SWIFTJ0250.7+4142 & 42.669 & 41.672 & 0.0145 & 43.18 & 0.25 $\pm$ 0.04 & 10.96 $\pm$ 0.06 \\
		SWIFTJ0251.6-1639 & 42.918 & -16.651 & 0.0110 & 42.96 & 0.39 $\pm$ 0.052 & 10.40 $\pm$ 0.19 \\
		SWIFTJ0304.1-0108 & 45.955 & -1.104 & 0.0136 & 43.68 & -0.61 $\pm$ 0.04 & 10.50 $\pm$ 0.05 \\
		SWIFTJ0331.4-0510 & 52.846 & -5.142 & 0.0128 & 42.62 & -0.70 $\pm$ 0.05 & 10.43 $\pm$ 0.06 \\
		SWIFTJ0333.6-3607 & 53.402 & -36.140 & 0.005 & 42.60 & 1.355 $\pm$ 0.013 & 10.78 $\pm$ 0.09 \\
		SWIFTJ0342.0-2115 & 55.516 & -21.244 & 0.0145 & 43.33 & 0.18 $\pm$ 0.03 & 10.84 $\pm$ 0.07 \\
		SWIFTJ0444.1+2813 & 71.038 & 28.217 & 0.0113 & 43.12 & 0.02 $\pm$ 0.03 & 10.069 $\pm$ 0.019 \\
		SWIFTJ0501.9-3239 & 79.899 & -32.658 & 0.0125 & 43.23 & 0.21 $\pm$ 0.04 & 10.24 $\pm$ 0.16 \\
		SWIFTJ0552.2-0727 & 88.047 & -7.456 & 0.008 & 43.63 & 0.31 $\pm$ 0.03 & 10.78 $\pm$ 0.09 \\
		SWIFTJ0601.9-8636 & 91.424 & -86.632 & 0.006 & 42.73 & 0.45 $\pm$ 0.02 & 10.55 $\pm$ 0.04 \\
		SWIFTJ0615.8+7101 & 93.902 & 71.037 & 0.0135 & 44.06 & 0.55 $\pm$ 0.04 & 10.68 $\pm$ 0.2 \\
		SWIFTJ0630.7+6342 & 98.197 & 63.674 & 0.0128 & 42.56 & 0.28 $\pm$ 0.02 & 10.26 $\pm$ 0.15 \\
		SWIFTJ0804.2+0507 & 121.024 & 5.114 & 0.0135 & 43.40 & 0.28 $\pm$ 0.05 & 9.93 $\pm$ 0.19 \\
		SWIFTJ0856.0+7812 & 133.907 & 78.223 & 0.0047 & 41.87 & -0.33 $\pm$ 0.03 & 10.93 $\pm$ 0.04 \\
		SWIFTJ0902.0+6007 & 135.493 & 60.152 & 0.0111 & 42.52 & 0.282 $\pm$ 0.019 & 9.82 $\pm$ 0.14 \\
		SWIFTJ0920.1+3712 & 139.992 & 37.191 & 0.0075 & 42.39 & 0.23 $\pm$ 0.03 & 10.50 $\pm$ 0.04 \\
		SWIFTJ0945.6-1420 & 146.425 & -14.326 & 0.0077 & 42.52 & 0.71 $\pm$ 0.03 & 10.60 $\pm$ 0.11 \\
		SWIFTJ0947.6-3057 & 146.917 & -30.949 & 0.0085 & 43.50 & -0.32 $\pm$ 0.16 & 9.67 $\pm$ 0.08 \\
		SWIFTJ0951.9-0649 & 147.979 & -6.823 & 0.0145 & 42.93 & 0.27 $\pm$ 0.03 & 10.84 $\pm$ 0.05 \\
		SWIFTJ0959.5-2248 & 149.873 & -22.826 & 0.008 & 43.29 & 0.02 $\pm$ 0.04 & 10.21 $\pm$ 0.03 \\
		SWIFTJ1001.7+5543 & 150.491 & 55.680 & 0.0037 & 42.53 & 1.01 $\pm$ 0.011 & 10.12 $\pm$ 0.08 \\
		SWIFTJ1005.9-2305 & 151.481 & -23.057 & 0.0128 & 42.74 & -0.19 $\pm$ 0.04 & 10.31 $\pm$ 0.11 \\
		SWIFTJ1023.5+1952 & 155.877 & 19.875 & 0.0039 & 42.55 & 0.39 $\pm$ 0.03 & 10.38 $\pm$ 0.04 \\
		SWIFTJ1031.7-3451 & 157.967 & -34.854 & 0.0107 & 43.48 & 0.76 $\pm$ 0.04 & 10.98 $\pm$ 0.09 \\
		SWIFTJ1048.4-2511 & 162.098 & -25.162 & 0.0125 & 43.08 & 0.33 $\pm$ 0.04 & 10.77 $\pm$ 0.05 \\
		SWIFTJ1106.5+7234 & 166.698 & 72.569 & 0.0088 & 43.29 & -0.05 $\pm$ 0.05 & 10.46 $\pm$ 0.13 \\
		SWIFTJ1132.7+5301 & 173.145 & 53.068 & 0.0033 & 41.47 & -0.90 $\pm$ 0.018 & 10.12 $\pm$ 0.04 \\
		SWIFTJ1139.0-3743 & 174.757 & -37.739 & 0.0097 & 43.58 & 0.074 $\pm$ 0.06 & 10.1 $\pm$ 0.3 \\
		SWIFTJ1139.8+3157 & 174.927 & 31.909 & 0.0089 & 42.45 & 0.20 $\pm$ 0.03 & 10.34 $\pm$ 0.11 \\
		SWIFTJ1143.7+7942 & 176.317 & 79.682 & 0.0065 & 42.38 & < -1.54 & 9.3 $\pm$ 0.3 \\
		SWIFTJ1203.0+4433 & 180.790 & 44.53 & 0.0023 & 41.65 & -0.073 $\pm$ 0.013 & 9.2 $\pm$ 0.2 \\
		SWIFTJ1206.2+5243 & 181.593 & 52.711 & 0.0028 & 42.09 & 0.844 $\pm$ 0.012 & 10.22 $\pm$ 0.11 \\
		SWIFTJ1209.4+4340 & 182.374 & 43.685 & 0.0030 & 41.79 & -0.65 $\pm$ 0.03 & 9.93 $\pm$ 0.06 \\
		SWIFTJ1210.5+3924 & 182.636 & 39.406 & 0.0033 & 43.10 & -0.10 $\pm$ 0.06 & 9.64 $\pm$ 0.09 \\
		SWIFTJ1212.9+0702 & 183.262 & 7.038 & 0.0070 & 42.44 & 0.610 $\pm$ 0.017 & 10.60 $\pm$ 0.05 \\
		SWIFTJ1217.3+0714 & 184.291 & 7.192 & 0.0080 & 42.64 & -0.64 $\pm$ 0.03 & 10.74 $\pm$ 0.08 \\
		SWIFTJ1218.5+2952 & 184.611 & 29.813 & 0.0129 & 42.90 & 0.74 $\pm$ 0.03 & 10.03 $\pm$ 0.05 \\
		SWIFTJ1219.4+4720 & 184.740 & 47.304 & 0.0017 & 41.11 & 0.05 $\pm$ 0.02 & 10.31 $\pm$ 0.03 \\
		SWIFTJ1225.8+1240 & 186.445 & 12.662 & 0.0084 & 43.70 & 0.18 $\pm$ 0.03 & 10.17 $\pm$ 0.08 \\
		SWIFTJ1235.6-3954 & 188.903 & -39.909 & 0.0118 & 43.96 & 0.49 $\pm$ 0.05 & 10.52 $\pm$ 0.16 \\
		SWIFTJ1252.3-1323 & 193.052 & -13.415 & 0.0146 & 42.81 & 0.40 $\pm$ 0.03 & 9.66 $\pm$ 0.07 \\
		SWIFTJ1304.3-0532 & 196.055 & -5.552 & 0.0040 & 41.89 & -0.544 $\pm$ 0.018 & 10.19 $\pm$ 0.06 \\
		SWIFTJ1304.3-1022 & 196.060 & -10.340 & 0.0104 & 42.82 & 0.307 $\pm$ 0.015 & 10.71 $\pm$ 0.06 \\
		SWIFTJ1306.4-4025A & 196.609 & -40.415 & 0.0150 & 43.24 & 1.02 $\pm$ 0.03 & 10.79 $\pm$ 0.14 \\
		SWIFTJ1313.6+3650B & 198.365 & 36.593 & 0.0029 & 41.01 & 0.55 $\pm$ 0.03 & 10.52 $\pm$ 0.03 \\
		SWIFTJ1325.4-4301 & 201.365 & -43.019 & 0.0018 & 43.02 & 0.37 $\pm$ 0.02 & 10.50 $\pm$ 0.15 \\
		SWIFTJ1333.5-3401 & 203.359 & -34.015 & 0.0130 & 42.95 & -0.37 $\pm$ 0.09 & 9.15 $\pm$ 0.11 \\
		SWIFTJ1335.8-3416 & 203.974 & -34.296 & 0.0077 & 42.91 & -0.41 $\pm$ 0.09 & 9.31 $\pm$ 0.13 \\
		SWIFTJ1341.9+3537 & 205.535 & 35.654 & 0.0035 & 41.48 & -1.15 $\pm$ 0.03 & 9.81 $\pm$ 0.04 \\
		SWIFTJ1345.5+4139 & 206.330 & 41.713 & 0.0086 & 42.46 & 0.19 $\pm$ 0.04 & 10.59 $\pm$ 0.08 \\
		SWIFTJ1351.5-1814 & 207.873 & -18.230 & 0.0122 & 42.72 & < -1.10 & 9.0 $\pm$ 0.3 \\
		SWIFTJ1413.2-0312 & 213.313 & -3.207 & 0.0062 & 43.30 & 0.24 $\pm$ 0.05 & 10.0 $\pm$ 0.2 \\
		SWIFTJ1442.5-1715 & 220.600 & -17.253 & 0.0093 & 43.36 & 0.709 $\pm$ 0.017 & 10.73 $\pm$ 0.08 \\
		\hline
	\end{tabular}
\end{table*}
	
\begin{table*}
	\centering
	\caption{The BAT AGN Sample cont.}
	\label{tab:BAT AGN2}
	\begin{tabular}{ l|c|c|c|c|c|r }
		\hline
		Swift-BAT ID & RA & Dec & Redshift & Log$_{10}$ L$_{14 - 195 \mathrm{keV}}$ (erg s$^{-1}$) & Log$_{10}$ SFR (M$_\odot$) & Log$_{10}$ M$_{\mathrm{Stellar}}$ (M$_\odot$ yr$^{-1}$) \\
		\hline
		SWIFTJ1515.0+4205 & 228.764 & 42.050 & 0.0086 & 42.54 & 0.67 $\pm$ 0.03 & 10.59 $\pm$ 0.05 \\
		SWIFTJ1717.1-6249 & 259.248 & -62.821 & 0.0037 & 42.51 & 0.23 $\pm$ 0.02 & 10.39 $\pm$ 0.10 \\
		SWIFTJ1838.4-6524 & 279.585 & -65.428 & 0.0133 & 43.68 & 0.34 $\pm$ 0.04 & < 10.00 \\
		SWIFTJ1844.5-6221 & 281.225 & -62.365 & 0.0142 & 43.25 & 0.08 $\pm$ 0.11 & < 10.26 \\
		SWIFTJ1937.5-0613 & 294.388 & -6.218 & 0.0103 & 42.76 & 0.45 $\pm$ 0.03 & 10.26 $\pm$ 0.14 \\
		SWIFTJ1942.6-1024 & 295.670 & -10.323 & 0.0058 & 42.75 & 0.28 $\pm$ 0.02 & 10.63 $\pm$ 0.07 \\
		SWIFTJ2009.0-6103 & 302.195 & -61.100 & 0.0149 & 43.40 & 0.41 $\pm$ 0.03 & < 10.15 \\
		SWIFTJ2052.0-5704 & 313.010 & -57.069 & 0.0114 & 43.41 & 0.55 $\pm$ 0.09 & 10.74 $\pm$ 0.04 \\
		SWIFTJ2201.9-3152 & 330.508 & -31.870 & 0.0087 & 43.32 & 0.51 $\pm$ 0.02 & 10.54 $\pm$ 0.12 \\
		SWIFTJ2209.4-4711 & 332.318 & -47.167 & 0.0058 & 42.49 & -0.15 $\pm$ 0.02 & 10.88 $\pm$ 0.09 \\
		SWIFTJ2302.1+1557 & 345.504 & 15.965 & 0.0066 & 42.11 & 0.07 $\pm$ 0.02 & 9.77 $\pm$ 0.04 \\
		SWIFTJ2304.9+1220 & 346.231 & 12.323 & 0.0079 & 42.50 & 0.92 $\pm$ 0.03 & 10.48 $\pm$ 0.02 \\
		SWIFTJ2318.4-4223 & 349.598 & -42.371 & 0.0052 & 43.29 & 1.011 $\pm$ 0.017 & 10.56 $\pm$ 0.11 \\
		\hline
	\end{tabular}
\end{table*}


\bsp	
\label{lastpage}
\end{document}